\documentclass[%
 reprint,
amsmath,amssymb,
aps,
]{revtex4-2}

\usepackage[utf8]{inputenc}

\usepackage{graphicx}% Include figure files
\usepackage{dcolumn}% Align table columns on decimal point
\usepackage{bm}% bold math
\usepackage{siunitx}
\usepackage{chemformula}
\usepackage{txfonts}
\usepackage{amsmath}
\usepackage{float}
\usepackage{hyperref}% add hypertext capabilities
\usepackage{changes}
\begin{document}

\preprint{APS/123-QED}

%\linenumbers

\title{High-fidelity single-spin shuttling in silicon}% Force line breaks with \\

\author{M. De Smet$^{1 \dag}$}
\author{Y. Matsumoto$^{1 \dag}$}
\author{A.M.J. Zwerver$^{1}$}
\author{L. Tryputen$^{2}$}
\author{S.L. de Snoo$^{1}$}
\author{S.V. Amitonov$^{2}$}
\author{S.R. Katiraee-Far$^{1}$}
\author{A. Sammak$^{2}$}
\author{N. Samkharadze$^{2}$}
\author{\"O. G\"ul$^{2}$}
\author{R. N. M. Wasserman$^{2}$}
\author{E. Greplov\'a$^{1}$}
\author{M. Rimbach-Russ$^{1}$}
\author{G. Scappucci$^{1}$}
\author{L.M.K. Vandersypen$^{1}$}
\email{L.M.K.Vandersypen@tudelft.nl}

\affiliation{$^{1}$QuTech and Kavli Institute of Nanoscience, Delft University of Technology, Lorentzweg 1, 2628 CJ Delft, The Netherlands \\ $^{2}$QuTech and Netherlands Organization for Applied Scientific Research (TNO), Stieltjesweg 1, 2628 CK Delft, Netherlands
}

\affiliation{$^{\dag}$ These authors contributed equally}

\date{\today}% It is always \today, today,

\begin{abstract}

The computational power and fault-tolerance of future large-scale quantum processors derive in large part from the connectivity between the qubits. One approach to increase connectivity is to engineer qubit-qubit interactions at a distance.~\cite{wallraff_strong_2004,dijkema_cavity-mediated_2025}.
Alternatively, the connectivity can be increased by physically displacing the qubits. This has been explored in trapped-ion experiments ~\cite{pino_demonstration_2021,sterk_closed_loop_2022} and using neutral atoms trapped with optical tweezers ~\cite{bluvstein_quantum_2022, bluvstein_logical_2023}. For semiconductor spin qubits, several studies have investigated spin coherent shuttling of individual electrons ~\cite{langrock_blueprint_2023,krzywda_interplay_2021,flentje_coherent_2017,fujita_coherent_2017,yoneda_coherent_2021,noiri_shuttling-based_2022,van_riggelen-doelman_coherent_2024,mortemousque_enhanced_2021,struck_spin-epr-pair_2024, jadot_distant_2021}, but high-fidelity transport over extended distances remains to be demonstrated. Here we report shuttling of an electron inside an isotopically purified Si/SiGe heterostructure using electric gate potentials. In a first set of experiments, we form static quantum dots, and study how spin coherence decays as we repeatedly move a single electron between up to five dots. Next, we create a traveling wave potential, formed with either one or two sets of sine waves, to transport an electron in a moving quantum dot. This second method shows substantially better spin coherence than the first. It allows us to displace an electron over an effective distance of \SI{10}{\micro \meter} in under 200 ns with an average fidelity of $99.5\%$. These results will guide future efforts to realize large-scale semiconductor quantum processors, making use of electron shuttling both within and between qubit arrays.

%~\cite{taylor_fault-tolerant_2005,vandersypen_interfacing_2017, li_crossbar_2018, boter_spiderweb_2022, jnane_multicore_2022, kunne_spinbus_2024}.

\end{abstract}

\maketitle

%\section{\label{sec:Introduction}Introduction}
%Quantum computing shows promise for tackling computationally challenging problems beyond the capabilities of classical computers.
To harness the full potential of quantum computation, errors will need to be corrected faster than they appear. The requirements for fault-tolerant quantum computation in terms of redundancy and error rates are generally eased when the connectivity between the qubits is stronger~\cite{bravyi_high-threshold_2024, xu_constant-overhead_2024}. Even then, the needed redundancy will quickly bring the number of qubits into the millions. Given the challenges in realizing large-scale monolithic quantum registers, approaches based on networks of spatially separated qubit registers that are connected by quantum links have gained traction.

Among the numerous quantum computing platforms, gate-defined semiconductor spin qubits~\cite{vandersypen_quantum_2019} have garnered significant attention. Recent advances in this field showcase extended spin coherence ~\cite{veldhorst_addressable_2014}, high-fidelity single-~\cite{yoneda_quantum-dot_2018, yang_silicon_2019,lawrie_simultaneous_2023} and two-qubit~\cite{xue_quantum_2022, noiri_fast_2022, mills_two-qubit_2022, tanttu_assessment_2024} gates, high-temperature operation ~\cite{undseth_hotter_2023, huang_high-fidelity_2024}, and universal control over up to six qubits ~\cite{philips_universal_2022}. Moreover, spin qubits are an attractive choice for densely packed quantum processors because of their compatibility with existing semiconductor fabrication techniques ~\cite{maurand_cmos_2016,zwerver_qubits_2022} and pitch of about 100 nm. 

Whereas the conventional two-qubit gate relies on the exchange interaction between spins in adjacent quantum dots, several avenues for increasing the connectivity between distant spin qubits on the same chip have been explored. Much effort has gone into engineering hybrid devices where superconducting resonators are used to couple electron or hole spins in quantum dots. This culminated in the recent observation of iSWAP oscillations between two spins separated by a few hundred micrometers ~\cite{dijkema_cavity-mediated_2025}. A promising alternative for distances of up to about ten micrometers consists in transporting (shuttling) spins across the chip, which can increase the connectivity within a qubit register and form a coherent link between qubit registers~\cite{taylor_fault-tolerant_2005,vandersypen_interfacing_2017, li_crossbar_2018,boter_spiderweb_2022, jnane_multicore_2022, kunne_spinbus_2024}. 

Two distinct procedures for spin shuttling exist, referred to as bucket-brigade and conveyor-mode. Bucket-brigade (BB) shuttling involves transporting a spin through an array of quantum dots by successively adjusting their electrochemical potentials. Successful charge transfer was realized across nine dots~\cite{mills_shuttling_2019} and spin-flip probabilities per hop below 0.01$\%$ were observed in both GaAs~\cite{baart_single-spin_2016} and Si/SiGe quantum dot arrays~\cite{zwerver_shuttling_2023, noiri_shuttling-based_2022}. Preservation of the spin phase was probed qualitatively in GaAs quantum dots~\cite{fujita_coherent_2017,flentje_coherent_2017} and quantitatively in a silicon double quantum dot, with phase-flip probabilities of $0.7\%$ ~\cite{yoneda_coherent_2021} and $0.1\%$ ~\cite{noiri_shuttling-based_2022} per hop during a Hahn-echo sequence. Near zero magnetic field, even lower phase-flip probabilities have been reported ~\cite{foster_dephasing_2024}. Additionally, it was shown in Ge quantum dots that diabatic BB shuttling in the presence of spin-orbit interaction can be used to generate single-qubit gates with fidelities of 99.97$\%$ ~\cite{wang_operating_2024}. Conveyor-mode shuttling is a technique in which a traveling wave potential transports the spin inside a moving quantum dot.  The traveling wave potential can be generated by a surface acoustic wave or by phase-shifted sinusoidal signals applied to successive gate electrodes. Also with conveyor-mode shuttling, charge transfer was demonstrated along channels of on the order of \SI{10}{\micro \meter} long~\cite{mcneil_-demand_2011,hermelin_electrons_2011, xue_sisige_2024}. Coherent spin transfer was studied by moving one of two electrons prepared in a spin singlet state~\cite{jadot_distant_2021,struck_spin-epr-pair_2024}. However, the relative performance of the two methods has not been compared directly so far. More importantly, coherent spin transfer over extended distances that is both fast and high-fidelity remains to be demonstrated.

In this work, we quantitatively investigate the phase flip probability when repeatedly moving an electron back and forth through a linear device in isotopically purified Si/SiGe, using Ramsey-style and Hahn-echo-style measurements. We compare the performance of bucket-brigade and conveyor-mode shuttling, including a conveyor-mode implementation that introduces sine waves with two frequencies instead of one, and study the performance as a function of the driving amplitude and shuttling speed. The conventional single-tone  conveyor reaches a shuttle speed of \SI{36}{m/s}, while the two-tone conveyor is still successfully operated at \SI{64}{m/s}. Finally, we execute interleaved RB to quantify the shuttling fidelity for a cumulative displacement of ten micrometers through the device, yielding a total transfer fidelity of $99.54 \pm 0.05\%$.

% ###
\begin{figure*}[t]
\includegraphics[width=\textwidth]{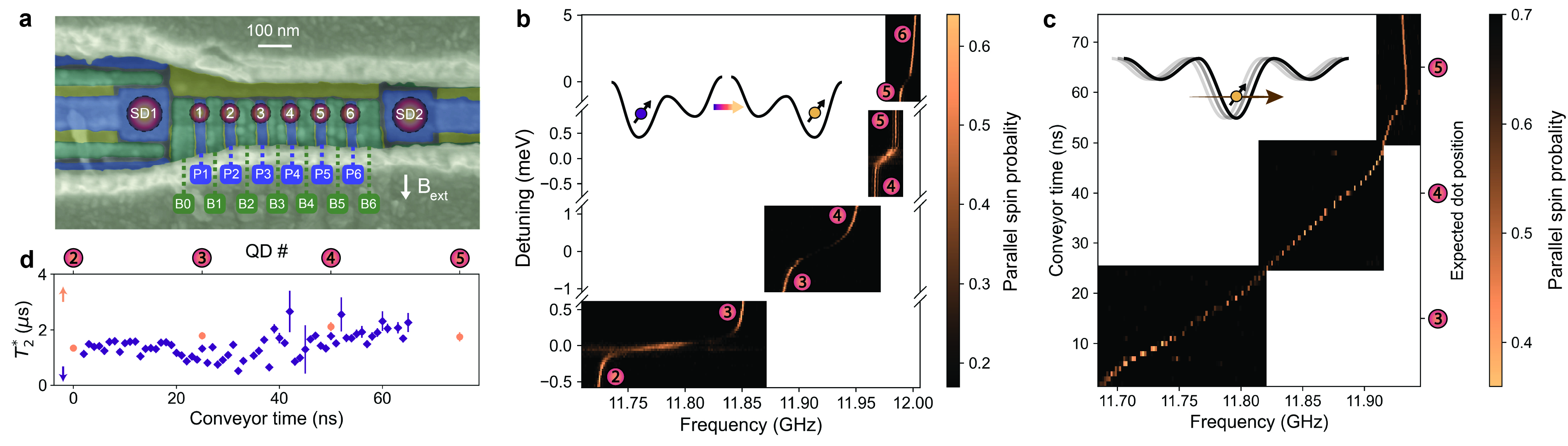}
\caption{\label{fig:fig1} \textbf{Device and characterization.}
a) False-colored scanning electron microscope (SEM) image of a nominally identical device to the one used in this work. The colors indicate different metallization layers. Six plunger (P$_i$ in blue), seven barrier (B$_i$ in green) and two screening gates (yellow) form a linear array of six quantum dots (indicated by numbered circles). Two sensing dots (SD$_i$) are placed at both ends of the array. A cobalt micromagnet, shown in sage gray, is placed on top of the active area. b) EDSR spectroscopy of the spin resonance frequency as a function of interdot detuning for each pair of neighbouring dots. A single spin is shuttled in bucket-brigade mode from dot 2 to the respective dot pair, a microwave burst is applied at a given interdot detuning and the spin is read out after shuttling back to dot 2. The color scale shows the spin flip probability as a function of the applied frequency and interdot detuning. The inset depicts bucket-brigade mode tunneling, transferring a single electron between dots. c) EDSR spectroscopy of the spin resonance frequency when shuttling a single spin in (two-tone, see below) conveyor-mode. The spin is initialized in dot 2, displaced by the conveyor potential by a distance controlled by the conveyor time, subjected to a microwave burst, displaced back to dot 2 and read out.  The color scale shows the spin flip probability as a function of the applied frequency and conveyor time (the expected conveyor displacement is shown on the right axis relative to the plunger gate positions). The inset depicts the traveling-wave potential that smoothly transfers the electron. d) Spin dephasing time in the gate-defined quantum dots (orange) and in a static two-tone conveyor at different conveyor times, corresponding to different locations along the channel (purple). In both cases, fits were performed with Gaussian decay and the error bars correspond to one standard deviation to
the fit.}
\end{figure*}

%\section{\label{sec:Results}Results}

\begin{figure*}[t]
\includegraphics[width=\textwidth]{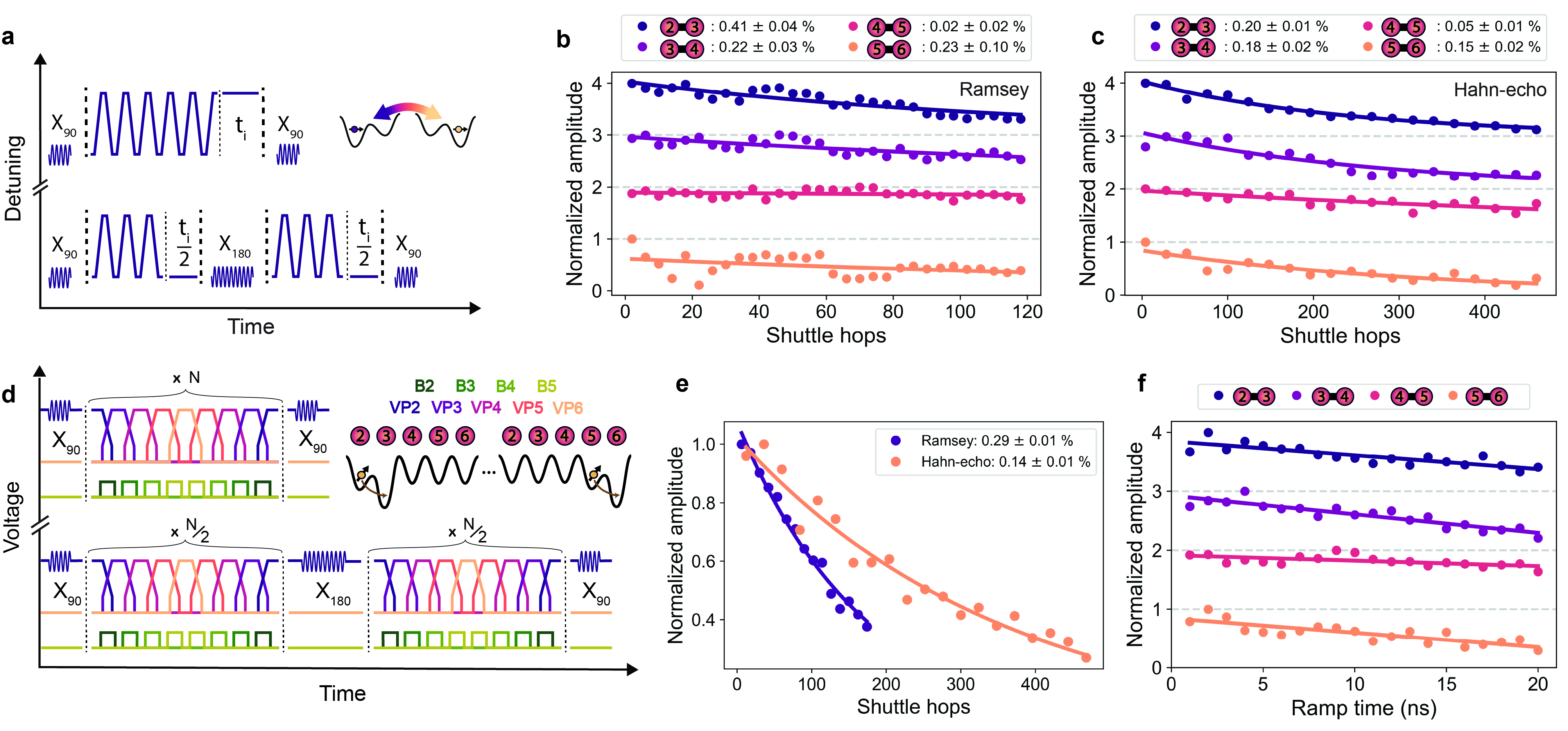}
\caption{\label{fig:fig2} \textbf{Bucket-brigade shuttling}.
a) Ramsey (top) and Hahn-echo (bottom) pulse sequences when shuttling repeatedly between sites of a double quantum dot. The detuning is pulsed roughly between the charge symmetric points of the (1,0) and (0,1) states of the double dot. X$_{90}$ and X$_{180}$ gates are applied in dot 2. b),c) Decay of the (b) Ramsey and (c) Hahn-echo fringe amplitude for each double dot with increasing number of shuttle hops, following the pulse scheme of panel a). Plots are normalized by the amplitude of the highest data point and are offset by 1 for clarity. A fitted exponential decay yields the indicated phase-flip probabilities. Performing the idling operation in different dots for the Ramsey and Hahn-echo sequences has no significant impact on the extracted phase-flip probability compared to the one standard deviation error bar. (see Supplementary Information). d) Ramsey (top) and Hahn-echo (bottom) pulse sequences when shuttling repeatedly through the array from quantum dot 2 to 6. e) Decay of the Ramsey and Hahn-echo fringe amplitude with increasing number of shuttle hops, following the pulse scheme of panel d) between dots 2 and 5. f) Normalized Hahn-echo fringe amplitude after shuttling forth and back twice through a double dot as a function of the ramp time. The uncertainties correspond to one standard deviation extracted from the fitting procedure.}
\end{figure*}

\subsection{\label{subsec:Device}Device design and characterization}

The device is fabricated on a $^{28}$Si/SiGe heterostructure, hosting a linear array of six quantum dots (Fig. ~\ref{fig:fig1}a). A cobalt micromagnet is deposited on top of the gate electrodes. Its stray field enables electric-dipole spin resonance (EDSR) for single-qubit rotations ~\cite{obata_coherent_2010} and separates the spin resonance frequencies for electrons in different dots. 

We place a reference electron in dot 1, and initialize an electron spin for shuttling in dot 2. All other quantum dots are emptied in order to investigate coherent spin shuttling through the array. For bucket-brigade, we pulse not only the electrochemical potentials but also raise the interdot barrier gate voltages one after the other to temporarily establish a large interdot tunnel coupling $t_c$ (between 17 and 55 GHz, see Supplementary Information). A large tunnel coupling is required for a rapid transfer between dots while maintaining adiabaticity ~\cite{flentje_coherent_2017, fujita_coherent_2017, baart_single-spin_2016}. Figure \ref{fig:fig1}b shows how the qubit resonance frequency abruptly shifts as an electron is displaced between neighbouring dots. For conveyor-mode shuttling we employ a number of phase-shifted sine signals applied to the plunger and barrier gates. The resulting traveling-wave potential transports the electron within a single potential minimum. This technique allows continuous control of the electron position along the array. Figure \ref{fig:fig1}c shows a continuous trend of the resonance frequency (albeit different than expected, see Supplementary Information). 

The local spin dephasing time $T_2^*$ can be probed for an electron in a static conveyor potential minimum as well as in any of the predefined dots (Fig. \ref{fig:fig1}d). In both cases, microwave driving is applied when the spin is in dot 2, while the idling time is spent in either a predefined dot or in a static conveyor potential minimum. $T_2^*$ shows an increasing trend towards dot 5, where the local magnetic field gradient from the micromagnet is weaker. Presumably charge noise leads to dephasing as it modulates the electron position in the gradient magnetic field, with additional dephasing contributions from residual hyperfine noise. We note that the dephasing times in the predefined dots are in general slightly longer than those in the conveyor minimum, indicating a tighter confinement hence reduced electrical susceptibility in the predefined dots.

\subsection{\label{subsec:BB}Bucket-brigade operation}

We study the bucket-brigade shuttling performance via Ramsey- and Hahn-echo-style measurements. We first investigate shuttling between adjacent dots (similar to \cite{yoneda_coherent_2021, noiri_shuttling-based_2022}) and next perform shuttling across multiple quantum dots. 

After initializing the reference spin in dot 1 and the spin in dot 2 in a parallel state, an $X_{\pi/2}$ gate is applied to the spin in dot 2. Next this spin is transported to the target double dot after which it is repeatedly transferred back and forth between the two sites (with a ramp time of 2 ns and a wait time of 1 ns in each dot, see the pulse schematics in Figure \ref{fig:fig2}a). The number of shuttle hops $N$ is varied while the idle time $t_i$ at the end of the shuttle sequence is adjusted, keeping the total sequence time constant. This allows to isolate the impact of hopping on the spin coherence. In the Hahn-echo measurements, a refocusing $X_{\pi}$ gate is applied in dot 2 after half the number of hops. After shuttling, a second $X_{\pi/2}$ rotation is applied in dot 2 before the spin state is measured using parity readout. 

Figures \ref{fig:fig2}b and \ref{fig:fig2}c show the decaying normalized amplitude of Ramsey and Hahn-echo fringes, respectively, with increasing number of shuttle hops. The fringes are measured in a rotating frame which is detuned from the qubit frequency by about 30 MHz. For all four double dot pairs, the decay is fitted to an exponential function $a \, (1-\epsilon)^n$, where $n$ is the number of shuttle hops and $\epsilon$ is a coherence loss per hop. We here define phase-flip probability $\frac{\epsilon}{2}$ as the probability for the spin to undergo a phase-flip error during shuttling, excluding initialization and measurement errors. This yields phase-flip probabilities below 0.5\% per hop (dividing the total error by the number of hops), and in some cases as low as 0.02\%. The phase-flip probability is lower in double dots with a smaller Zeeman splitting difference (see Fig.\ref{fig:fig1}b), which agrees with numerical simulations (see Supplementary Information). Double dot 5-6 does not follow this trend, which is possibly related to the comparatively large voltage needed on gate B5.

Next we evaluate spin coherence during sequential shuttling through the entire array. We pulse both the interdot detuning and the tunnel coupling between the successive double dots in turn. Pulsing the tunnel couplings enables us to reach a high tunnel coupling between the target dots to ensure adiabatic transport, while suppressing charge leakage to adjacent dots by reducing the corresponding tunnel couplings (Fig. \ref{fig:fig2}d). Figure \ref{fig:fig2}e shows the decaying amplitude of Ramsey and Hahn-echo fringes, respectively, with increasing number of shuttle hops between dot 2 to dot 5 (BB shuttling works best between these dots, see Supplementary Information). In contrast to the shuttling benchmark in double dots (Figs. \ref{fig:fig2}b and c), we vary the number of hops without additional idle times at the end of the shuttling sequence. This means spin coherence lost during the shuttling process or while idling in individual dots cannot be distinguished. However, this approach does reveal how spin coherence is affected overall when shuttling under realistic conditions across many quantum dots. We obtain an average phase-flip probability per hop of 0.29  $\pm$ 0.01\% with the Ramsey protocol and of 0.14 $\pm$ 0.01\% with the Hahn-echo protocol.

The spin dephasing time $T_2^*$ extracted from the BB shuttling measurements in Figure \ref{fig:fig2}e is \SI{1.04}{\micro s}, compared to an average $T_2^*$ of \SI{1.75}{\micro s} in the static dots. 
This indicates that the act of hopping between dots increases the phase\added{-}flip probability. As stated in other works ~\cite{yoneda_coherent_2021, feng_control_2023, van_riggelen-doelman_coherent_2024}, charge noise couples more strongly to the qubit in the low detuning regime of each double quantum dot, where the resonance frequency is highly sensitive to detuning fluctuations (see Fig. 1b). Phase flips from hopping can also occur in case the charge transitions are not perfectly adiabatic with respect to the tunnel couplings. Small uncertainties in the timing of charge transfer then lead to dephasing due to the Larmor frequency difference between the quantum dots. Such diabatic transitions can in principle result either from the high-frequency components of the voltage ramp, or from rapid electric field fluctuations arising from charge noise. Figure \ref{fig:fig2}f shows that the phase\added{-}flip errors monotonously decrease with decreasing ramp time down to 2 ns. Furthermore, the loss of phase coherence increases with the Zeeman splitting difference. These observations are consistent both with enhanced dephasing halfway the interdot transition (see the simulations in the Supplementary Information) and with diabatic transitions from rapid electric field fluctuations as the phase\added{-}flip mechanism. 
Contrary to these trends, we systematically observe larger phase\added{-}flip probabilities with 1 ns ramps than with 2 ns ramps. Presumably the 1 ns ramp time in combination with the 1 ns wait time between ramps does not allow the pulse to reach full amplitude given the waveform generator bandwidth. 

\begin{figure*}

\includegraphics[width=0.9\textwidth]{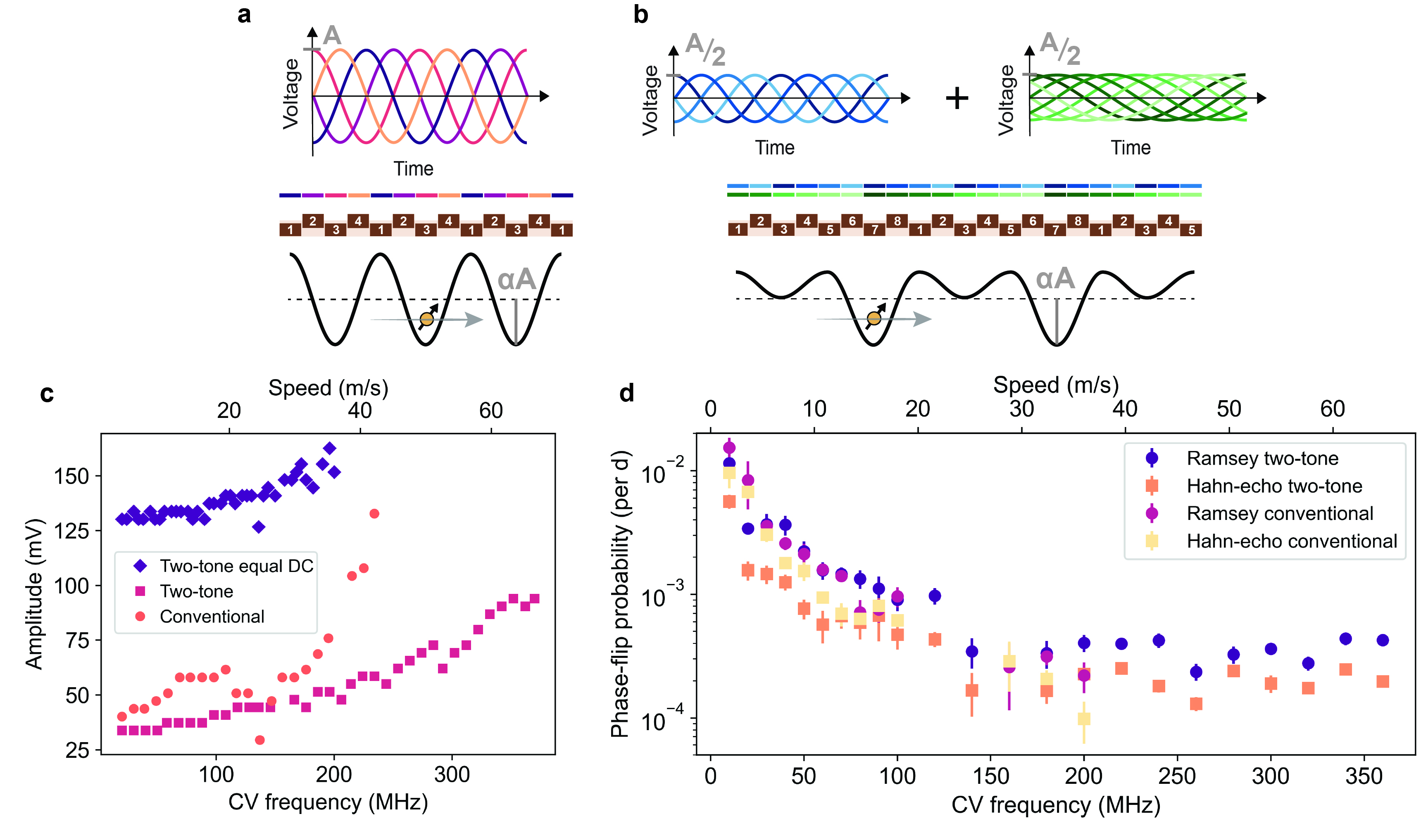}
\caption{\label{fig:fig3} \textbf{Conveyor-mode shuttling.} a) Oscillating gate voltages (top) and the associated traveling-wave potential in the quantum well (bottom) for the conventional four-phase conveyor. Gates can be divided into four sets (numbered 1-4) and the color above each gate corresponds to the phase of the applied sine signal. A indicates the amplitude of the voltage signal and the lever arm $\alpha$ expresses the conversion of gate voltage to dot potential. In the experiment, two different amplitudes are applied to gates in separate fabrication layers to compensate for the difference in lever arm (not shown in the figure for simplicity).  b) Gate voltages as a sum of two oscillating signals with frequencies $f$ and $f/2$ (top) and the associated traveling-wave potential in the quantum well (bottom) for the two-tone conveyor. Gates can be divided into eight sets (numbered 1-8) and the colors above each gate corresponds to the phases of the applied sine signals. Each sine wave uses half the amplitude of the four-phase conveyor. c) Minimal conveyor amplitude required for successful charge transfer by shuttling, accounting for the known attenuation in the transmission lines (see Suppl. Fig. 10 for a figure accounting for the filter function built-in to the AWG as well). The conveyor potential minimum is displaced from below P2 to below P5, a resonant microwave burst is applied to flip the spin using EDSR (if the electron was successfully transferred), the potential minimum is moved back to below P2, and the electron spin is read out. Individually adjusted DC voltages are used in the orange and pink cases (see Supplementary Information). For the purple data points, a common plunger and a common barrier DC voltage is applied. The shuttling speed is calculated using the applied conveyor frequency and the \SI{180}{\nano \meter} wavelength set by the gate pitch. d) Phase-flip probability per nominal $d=$ \SI{90}{\nano \meter} plunger-to-plunger distance for the conventional conveyor and the two-tone conveyor as a function of the applied conveyor frequency. All conveyor amplitudes are set to $A = 100$ mV, except for the two-tone conveyor amplitudes above 200 MHz, which are set to 120 mV. The need to increase the amplitude can be understood from panel (c). For conveyor frequencies between 100 MHz and 160 MHz, spin coherence in the conventional conveyor is completely lost (see the missing data points). While the underlying reasons for this phenomenon remain unclear, we provide additional comments in the Supplementary Information. The phase-flip probabilities are extracted from the exponential decay of Ramsey or Hahn-echo fringes with increasing shuttling distance. The error bars correspond to one standard deviation extracted from the fit.}
\end{figure*}

\subsection{\label{subsec:CV}Conveyor-mode operation}

Turning to conveyor-mode shuttling~\cite{langrock_blueprint_2023, seidler_conveyor-mode_2022, struck_spin-epr-pair_2024}, the traditional approach makes use of four phase-shifted voltage signals applied to a set of gate electrodes, with the voltage 
\begin{equation}
\label{eq:4phase_signals}
V_n (t) = V_n^{DC} - A \sin{\Big(2\pi f t - \phi_n} \Big)
\end{equation}
applied to gate $n$, where $\phi_n = \phi' + (n \mod{4}) \, \pi/2$ and $\phi'$ is a phase offset that determines the initial position of the potential minima. Here, $V^{DC}_n$ represents a DC voltage offset and $f$ is the conveyor frequency. In most measurements shown below, we individually adjust the DC voltage applied to each gate. The voltage signals and resulting traveling potential wave are illustrated in Figure \ref{fig:fig3}a.

We additionally propose and assess a novel two-tone conveyor approach with voltage signals given by 
\begin{multline}
\label{eq:8phase_signals}
V_n (t) = V_n^{DC} - \frac{A}{2} \Biggr[ \sin{\Big(2\pi f t - \phi_n} \Big) + \sin{\Big(\pi f t - \theta_n \Big)} \Biggr].
\end{multline}
This conveyor incorporates a second sine wave with exactly half the original conveyor frequency and $\theta_n = \phi'/2 + (n+1 \mod 8) \, \pi/4$. While requiring twice the number of distinct control signals compared to the conventional conveyor, a significant advantage in the shape of the traveling potential wave (Figure \ref{fig:fig3}b) is achieved. Destructive interference at every second potential minimum strongly suppresses charge leakage to neighboring moving dots during shuttling. This is especially relevant in the presence of disorder in the background potential landscape. In Eqs.~\ref{eq:4phase_signals} and~\ref{eq:8phase_signals}, we have assumed that equal amplitude sine signals are applied to all gates. Given that the barrier gates are less well coupled to the channel than the plunger gates, the amplitude applied to the barrier gates was 1.4 times that applied to the plunger gates, a ratio which we found works well (all values below refer to the plunger gate amplitudes). We note that also the two tones could be applied with different amplitudes.

To determine the amplitude required for successful charge transfer by the conveyor, we shuttle a single spin from below gate P2 to close to below gate P5, apply an $X_{\pi}$ gate, and shuttle back. We then identify the minimal conveyor amplitude for which the spin flip from the $X_{\pi}$ gate is detected at readout, indicating successful charge transfer (the transition typically occurs within a few mV). Figure \ref{fig:fig3}c shows the minimal amplitude as a function of the main conveyor frequency, where we observe that the conventional conveyor necessitates a higher amplitude than the two-tone conveyor. The data suggests that even faster transport would be feasible, especially for the two-tone conveyor (the shuttling speed was limited by the output filter of the control hardware). For comparison, we also show the minimum amplitude for the two-tone conveyor when the same DC voltage is applied to all plunger gates, and another fixed DC voltage is applied to all barrier gates ~\cite{seidler_conveyor-mode_2022, struck_spin-epr-pair_2024, xue_sisige_2024}. Operating a conveyor with equal DC voltages applied to all gates would reduce the overhead involved in operating long-distance shuttling channels. We see that when the disorder in the background potential landscape is not compensated for by the local DC voltages, we can still successfully displace charges using the conveyor, albeit requiring a higher conveyor amplitude.

We characterize coherent spin transport in conveyor-mode by recording the phase-flip probability in Ramsey- and Hahn-echo-style sequences, using similar methods as used for bucket-brigade shuttling. Figure \ref{fig:fig3}d shows the phase-flip probability, defined per plunger-to-plunger distance $d$ to allow comparison with BB, for both conveyor types. Initially the probability decreases with increasing speed. In general, faster transfer means the spin has less time to dephase while it is transported in the moving dot. This behavior follows the prediction by ~\cite{langrock_blueprint_2023}. Above 150 MHz, the measured phase-flip probability does not keep decreasing but saturates, for reasons that are not fully understood. The Ramsey phase flip probabilities are similar between the conventional and two-tone conveyor, whereas the echo sequences show better results for the two-tone version. Importantly, the two-tone case allows for significantly higher speeds with modest conveyor amplitudes (see Fig.~\ref{fig:fig3}c). Even though the observed phase-flip probabilities level off above 150 MHz, faster shuttling is still advantageous in future scenarios where a subset of qubits is subject to decoherence while others are shuttled.
% \added{A strong increase in phase-flip probability due to spin-valley excitations, predicted by ~\cite{langrock_blueprint_2023}, is however not observed in our shuttle speed range.}

The data pertaining to dephasing in a moving conveyor is well-fitted to a single exponential, indicating high-frequency components in the noise experienced by the spin. For the static (two-tone) conveyor, the decay is close to Gaussian, pointing at low-frequency noise dominating the decay. This suggests that the exponential decay in a moving conveyor results primarily from moving through a spatially varying but quasi-static noise environment. Moreover, the extracted (exponential) dephasing time for both the conventional and two-tone conveyor during shuttling is longer than the (Gaussian) decay time in the static two-tone conveyor case in Figure ~\ref{fig:fig1}d (see Supplementary Information). This indicates that shuttling in conveyor mode does not suffer from important additional dephasing channels and in fact may benefit from motional narrowing effects~\cite{langrock_blueprint_2023}. We note there is no clear dependence of $T_2^*$ on conveyor speed, which would suggest that a motional narrowing effect in this finite-size device is already fully present at the slowest shuttling speeds, giving rise to an increase in dephasing times.

\begin{figure*}[]
\includegraphics[width=0.8\textwidth]{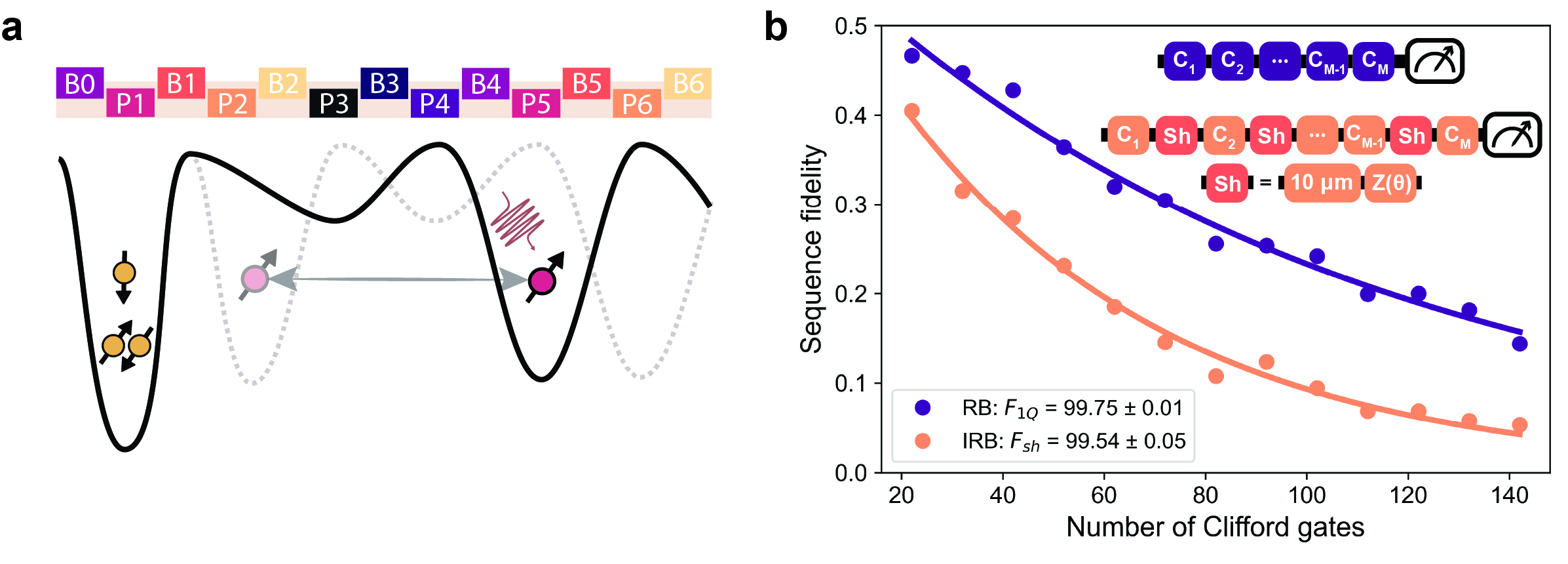}
\caption{\label{fig:fig4}\textbf{Shuttling fidelity.}
a) Schematic of the potential landscape when performing interleaved randomized benchmarking for repeated shuttling of a single spin using a two-tone conveyor. Quantum dot 1 hosts three electron spins serving as an ancilla for PSB readout. Specifically, we use a 300 MHz conveyor frequency, displace the electron for 4 ns and move it back in 4 ns, and repeat this over 24 rounds, corresponding to a cumulative distance of \SI{10}{\micro \meter}, which takes 184 ns. Single-qubit control with microwave pulses is always performed with the electron roughly under gate P5.  b) Single-qubit RB (purple) as a reference for the shuttling gate interleaved RB (IRB) (orange). Here, the uncertainties in the gate fidelities represent one standard deviation and are obtained by bootstrap resampling \cite{barends_superconducting_2014}. The inset shows the circuit for the single-qubit reference (purple) and \SI{10}{\micro \meter} shuttling interleaved RB (orange), where $C_i$ are single-qubit Clifford gates. The shuttling operation consists of moving the electron back and forth over an accumulated \SI{10}{\micro \meter} distance, and a virtual Z phase compensation. This randomized benchmarking data was acquired with the superconducting magnet in persistent mode, contrary to the measurements presented in all previous figures. Shuttling interleaved RB data with the magnet in driven mode, along with Ramsey measurements, can be found in the Supplementary Information.
}
\end{figure*}

\subsection{\label{subsec:IRB}Shuttling fidelity}

Finally, we characterize the shuttling fidelity of the high-speed two-tone conveyor by employing randomized benchmarking (RB). Figure \ref{fig:fig4}a shows a schematic of the experiment. The spin is initialized in dot 2, then the conveyor potential is switched on and the electron is shuttled to below gate P5, where we execute Clifford operations using EDSR with the electron in the static conveyor minimum (in this way, we avoid potential errors in transferring the spin from a fixed dot potential to a conveyor minimum). To benchmark the shuttling fidelity, we compared the decay of a standard RB sequence with that of an interleaved RB sequence. We interleave repeated conveyor shuttling operations, roughly between gates P2 and P5, followed by a virtual Z operation that ensures that the total operation (ideally) corresponds to the identity gate. For this part of the experiment, we further optimize the pulsed offset voltages applied to the plunger and barrier gates during shuttling using a global optimizer ~\cite{hansen_cma_2006}. Additionally, we operate the magnet in persistent mode instead of in driven mode, which results in twice longer $T_2^*$ values. Figure \ref{fig:fig4}b shows the results for both the reference and interleaved RB sequence in this regime. We obtain a single-qubit gate fidelity $F_{1Q}$ of 99.75 $\pm$ 0.01$\%$ and a shuttling fidelity $F_{\text {sh}}$ of 99.54 $\pm$ 0.03$\%$ for shuttling over a distance of \SI{10}{\micro \meter} (Shuttling interleaved RB data taken under the same conditions as Fig. ~\ref{fig:fig3} can be found in the Supplementary Information). 

Our analysis indicates that the conveyor shuttling fidelity is mainly limited by the ratio between the shuttling time and the static $T_{2}^{*}$.

Additional improvements in the coherence time can be achieved by reducing charge noise and minimizing magnetic field gradients. Remarkably, the obtained $F_{\text {sh}}$ already corresponds to a shuttle fidelity of 99.996$\%$ over a plunger-to-plunger distance, and the fidelity decays by $1/e$ after about \SI{2.25}{\milli \meter}, if we simply extrapolate exponentially.

%\section{\label{sec:conclusion}Conclusion}
\subsection{\label{subsec:conclusion}Conclusion}

This work demonstrates coherent bucket-brigade shuttling in silicon across multiple dots. We obtain phase-flip probabilities below 0.2\% per hop when shuttling back and forth through the array. Conveyor-mode shuttling, using both conventional four-phase and novel two-tone conveyors, exhibits stable and highly coherent spin transport, with phase-flip probabilities about one order of magnitude lower than our best BB results for a comparable distance. Using a two-tone conveyor, we are able to shuttle an electron back and forth over a cumulative distance of \SI{10}{\micro \meter} in less than \SI{200}{\nano \s} and with a fidelity of 99.5\%.

In this comparative investigation of shuttling methods, we find that conveyor-mode shuttling allows faster spin transfer with a higher fidelity, limited by the inherent dephasing in the device. It avoids sequential adiabatic interdot crossings and repeated charge delocalization intrinsic to bucket-brigade shuttling. For successful conveyor-mode operation, we must ensure the electron does not escape from the traveling potential minimum, which can be especially challenging in the presence of disorder. With only a limited increase in control parameters, the two-tone conveyor reduces this escape probability and we find it allows higher shuttling speeds with lower drive amplitudes. 

We note that future devices would likely not utilize a micromagnet on top of the shuttling channel, which could lead to even higher shuttling fidelities as the Zeeman splitting gradients will be orders of magnitude smaller. Moreover, one can encode the qubit in the $S$-$T^0$ subspace and shuttle both electrons sequentially to further protect the shuttled information against quasi-static noise~\cite{taylor_fault-tolerant_2005}. However, whereas the valley splitting in this device is estimated to be above 170 $\mu$eV (see Supplementary Information), the possibility of encountering local regions with low valley splitting increases for longer quantum links~\cite{volmer_mapping_2024}. This would require to (locally) slow down the conveyor~\cite{langrock_blueprint_2023} in order to avoid detrimental valley excitations, which would degrade the shuttling fidelity. Furthermore, we note that the best results are obtained when tweaking the individual DC gate voltages, which ideally is to be avoided in future implementations, requiring a lower level of background potential disorder ~\cite{neyens_probing_2024}. 

The shuttling fidelities needed for realizing fault-tolerant spin qubit architectures depend significantly on the level of connectivity and the target shuttling distances involved. Broadly speaking, though, for shuttling to be attractive to increase connectivity within a qubit register, shuttling error rates should be well below the two-qubit gate error rates. For establishing links between registers, higher error rates can be tolerated, especially if the local registers offer long-lived quantum memories ~\cite{nickerson_freely_2014}. In both scenarios, the shuttling fidelities achieved here show great promise.

\section*{\label{sec:ackn}Acknowledgments}
We wish to thank S.G.J. Philips for writing control libraries and designing the PCB, R. Schouten, R. Vermeulen, O. Benningshof and T. Orton for support with the measurement setup and dilution refrigerator, members of the Vandersypen, Veldhorst, Scappucci and Dobrovitski groups for fruitful discussions. We thank D. Michalak for managing the TEM study. We acknowledge financial support from the Army Research Office (ARO) under grant number W911NF-17-1-0274 and W911NF2310110 and the Dutch Ministry for Economic Affairs through the allowance for Topconsortia for Knowledge and Innovation (TKI). M. Rimbach-Russ acknowledges support from the Netherlands Organization of Scientific Research (NWO) under Veni Grant No. VI.Veni.212.223. E. Greplov\'a acknowledges the project Engineered Topological Quantum Networks (Project No.VI.Veni.212.278) of the research program NWO Talent Programme Veni Science domain 2021 which is financed by the Dutch Research Council (NWO). The views and conclusions contained in this document are those of the authors and should not be interpreted as representing the official policies, either expressed or implied, of the ARO or the US Government. The US Government is authorized to reproduce and distribute reprints for government purposes notwithstanding any copyright notation herein. 
% \end{acknowledgments}

\section*{\label{sec:contrib}Author contributions}
M.D.S. and Y.M. performed the experiments and data analysis. Simulations were carried out by Y.M. Device screening was done by A.M.J.Z., M.D.S, N.S., \"O.G., and R.N.S.W. A.M.J.Z. and M.D.S. conducted a first version of the experiment. Libraries for experimental control were written by S.L.S. and Y.M. M.D.S., Y.M., M.R.R. and L.M.K.V. contributed to data interpretation. L.T. fabricated the device, while S.V.A., L.T. and N.S. refined the device design. A.S. and G.S designed and grew the heterostructure. S.R.K-F. and E.G. implemented the global optimizer for the pulsed offset voltages of the conveyor. M.D.S., Y.M. and L.M.K.V. wrote the manuscript with comments by all authors. L.M.K.V. conceived and supervised the project.

\section*{\label{sec:interests}Competing interests}
M.D.S. and Y.M. are co-inventors of a patent application (NL2036142) concerning two-tone conveyor-mode shuttling. The other authors declare no competing interests.

\section*{\label{sec:contrib}Data availability}
The raw measurement data and the analysis supporting the findings of this work are available on a Zenodo repository (https://doi.org/10.5281/zenodo.10834811).

\bibliographystyle{naturemag}

\bibliography{manualbib}

\section*{\label{sec:methods}Methods}

The device is fabricated on an isotopically-purified $^{28}$Si/SiGe heterostructure featuring a \SI{7}{\nano \meter} quantum well. The confinement potentials are defined by three layers of Ti:Pd gates separated by \ch{Al2O3}, serving as screening, plunger (P) and barrier (B) gates, respectively. Sensing dots are placed at the ends of the linear array to facilitate charge sensing and to act as electron reservoirs. An external magnetic field of \SI{260}{\milli T} is applied in the plane of the quantum well and all experiments are performed in a dilution refrigerator set to a temperature of \SI{200}{\milli K} ~\cite{undseth_hotter_2023}.

We operate at the (3,1)-(4,0) charge transition of dots 1 and 2, creating a sizable readout window for parity Pauli-spin-blockade ~\cite{harvey-collard_high-fidelity_2018, philips_universal_2022}. Initialization of two electron spins is done by ramping from the (4,0) to the (3,1) charge state, subsequently performing parity read out of the spin states and post-selecting the single-shot runs with either the parallel or anti-parallel measurement outcomes. 

%%%%

 Spatial variation of the conduction band minimum can make that a challenging task. This disorder in the background potential can have various origins. First, local lattice strain induced by the deposition and cooldown of the gate electrodes gives rise to local modulations of the conduction band minimum. Second, charge defects in the gate dielectrics can lead to electrostatic disorder in the shuttling channel. Third, non-uniformity in the fabrication of the gate electrodes can lead to variations in the gate lever arms. 

%%%%

We analyze the shuttling operation fidelity using interleaved randomized benchmarking. For the reference sequence, we perform conventional randomized benchmarking based on randomly selected sequences of Clifford operations. We measure the probability for the final spin state to correspond to the initial spin-up or spin-down state as a function of the number of Clifford operations. The initial state is selected by an optional microwave burst before the RB sequence. The final step involves subtracting the measured probabilities for the two initial states from each other to minimize the uncertainty associated with the exponential fitting of the data. As the number of applied Clifford gates $N$ increases, the return probability decreases. We fit this decay to $F_{\mathrm{seq}} = A p_{\mathrm{c}}^N$, where $p_{\mathrm{c}}$ is the depolarizing parameter and the amplitude $A$ depends on state preparation and measurement errors. The average fidelity per Clifford operation $F_{\mathrm{c}}$ and per single primitive gate $F_{\mathrm{c}}^{\text {single }}$ are then calculated using 
\begin{equation}
F_{\mathrm{c}}=\left(1+p_{\mathrm{c}}\right) / 2,
\end{equation}
\begin{equation}
F_{\mathrm{c}}^{\text {single }}=\frac{1+p_{\mathrm{c}}^{\text {single }}}{2} \sim 1-\frac{1-F_{\mathrm{c}}}{1.875}
\end{equation}

We estimate the fidelity for coherently shuttling of a spin over about \SI{10}{\micro \meter} by interleaved randomized benchmarking. In between successive Clifford operations, the electron is shuttled back and forth repeatedly over a distance of \SI{10}{\micro \meter}. We call the depolarizing parameter for the interleaved sequence $p_{\text{sh}}$. The fidelity for shuttling over \SI{10}{\micro \meter} $F_{\text {sh}}$ is then estimated as 
$$
F_{\text {sh }}=\frac{1+\left(p_{\text {sh }} / p_{\mathrm{c}}\right)}{2} \;.
$$

The uncertainties in the gate and shuttling fidelities are estimated using a bootstrap resampling\cite{barends_superconducting_2014}. By repeating the resampling process $10^4$ times, we obtain distributions of the single-qubit gate fidelity $F_c^{\text{single}}$ and the shuttling fidelity $F_{sh}$. The final uncertainties are calculated as the standard deviations of these distributions.

The total effective shuttling distance $D_{sh}$ in Figure ~\ref{fig:fig4} is calculated as

\begin{equation}
D_{sh}=2N \, f_{CV} \, t_{sh} \, 2d
\end{equation}

Here $N$ is the number of shuttle rounds, $f_{CV}$ is the conveyor frequency, $t_{sh}$ is the one-way shuttle time. Note that $2d$ is the shuttled distance per conveyor period, which is double the plunger-to-plunger distance $d$. Given 24 shuttle rounds at a 300 MHz conveyor frequency, with $t_{sh}$ = \SI{4}{\nano \s} and $2d$ = \SI{180}{\nano \meter}, $D_{sh}$ yields \SI{10.368}{\micro \meter}. We note that local disorder in the potential landscape at the boundaries of the shuttle trajectory could be a limiting factor for the accuracy of this shuttle distance estimate.

% \appendix
\section*{\label{sec:Supp}Supplementary information}

\subsection{\label{supp:subsec:voltage_}Gate voltage conditions during shuttling}

The pulse sequence schematics for BB shuttling in Figure ~\ref{fig:fig2}d and for CV shuttling in Figure ~\ref{fig:fig3}a and b are simplified representations of the respective shuttling methods. Supplementary Fig. ~\ref{supp:fig:BB_CV_pulses} shows the actual voltage signals applied in the experiments for BB shuttling between dot 2 and 5 (Figure ~\ref{fig:fig2}e) and CV shuttling during IRB (Figure ~\ref{fig:fig4}), transporting the spin forth and back once. In bucket-brigade, we employ a 2 ns detuning ramp time, as we systematically observe larger phase-flip probabilities with 1 ns ramps than with 2 ns ramps. Presumably the 1 ns ramp time in combination with the short wait time between ramps does not allow the pulse to reach full amplitude given the waveform generator bandwidth.  After the electron has been shuttled through a quantum dot, the dot potential is farther detuned during the next transitions in order to prevent transferring the charge backwards. For the CV measurement, two-tone sinusoidal signals are applied with different amplitudes to gates in different metal layers in order to compensate for the uneven lever arms. Instead of P2, we use the virtual plunger VP2 to avoid changing the electron occupation in quantum dot 1. The spin is transported from approximately underneath gate P5 to P2 and back. To avoid activating the exchange interaction with the reference spin in dot 1, we shuttle up to a $\frac{\pi}{5}$ conveyor phase offset from under P2. The location of the spin under gate P5 is then determined by the integer 4 ns conveyor time at a frequency of 300 MHz. The DC voltage conditions are given in Table ~\ref{Supp:tab:DC_voltage}. In the conveyor case, these values include pulsed voltage offsets during the entire duration of conveyor operation, as we need different DC voltages while shuttling than during initialization and readout.

To determine the DC voltages for conveyor operation, we usually start by tuning the predefined static quantum dots. We then increase the DC voltages of the barrier gates, while lowering the DC voltages on the plunger gates, such that the potential landscape is flattened (starting from having six local potential minima forming six dots). We then track the spin resonance frequency during shuttling using EDSR spectroscopy. Once a smooth variation of the spin resonance frequency is reached (as opposed to the stair-case like variation for bucket-brigade shuttling), we further improve the shuttling fidelity by either manually or algorithmically fine-tuning the DC voltages to minimize the Ramsey phase-flip probability during back-and-forth shuttling.

\setcounter{figure}{0}
\renewcommand{\figurename}{FIG.}
\renewcommand{\thefigure}{S\arabic{figure}}

\begin{figure}[h]
\centering
\includegraphics[width=0.48\textwidth]{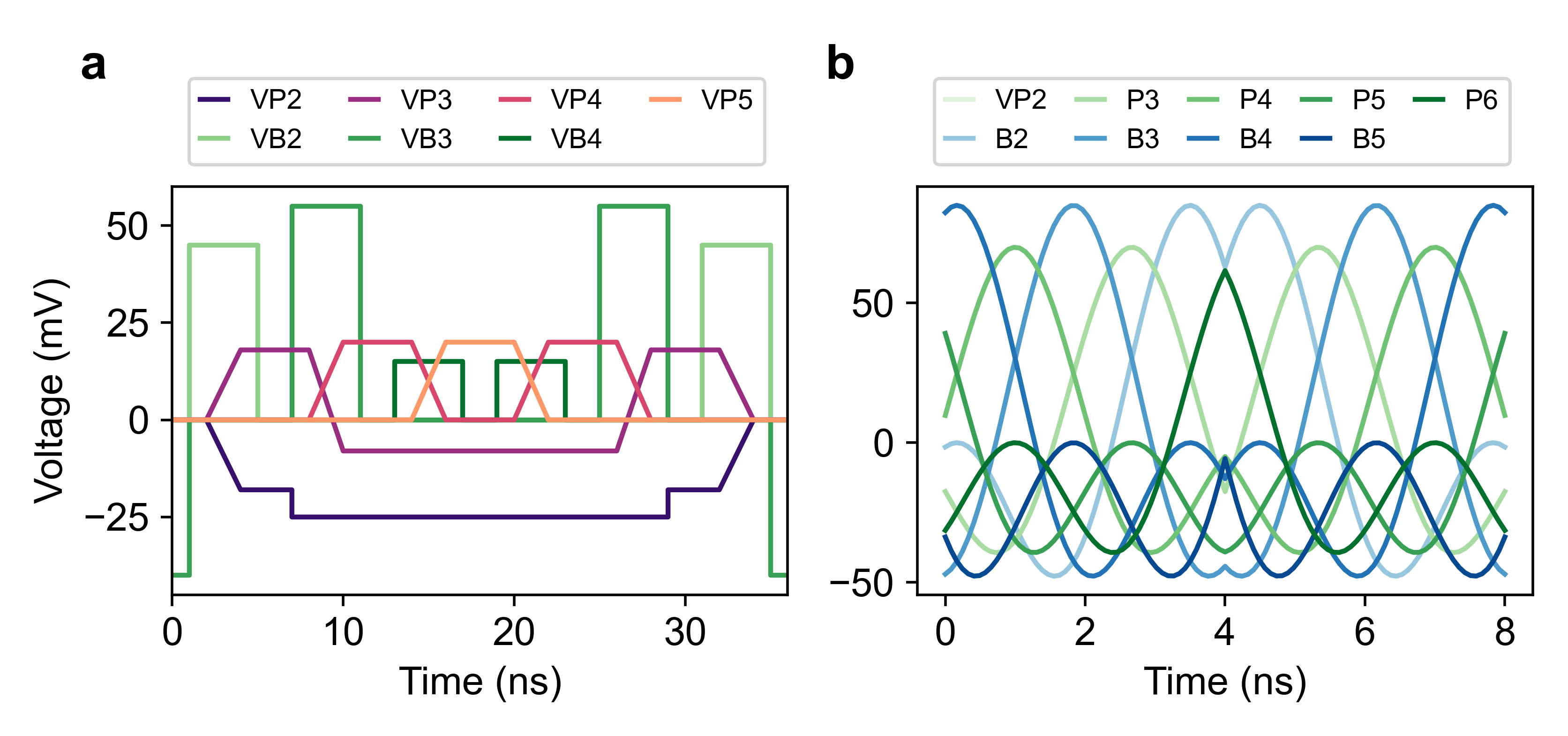}
\caption{\label{supp:fig:BB_CV_pulses} \textbf{Applied shuttling voltages.} a) Pulses for BB shuttling from dot 5 to 2 and back. b) Sinusoidal signals for two-tone CV shuttling from approximately under gate P2 to P5 and back at a main conveyor frequency of \SI{300}{MHz}.
}
\end{figure}

\renewcommand{\tablename}{TABLE}
\renewcommand{\thetable}{S\arabic{table}}

\begin{table}[h]
\centering
\caption{DC voltages applied to the shuttling channel gates during bucket-brigade (BB) and conveyor-mode (CV) shuttling.}
\label{Supp:tab:DC_voltage}
\begin{tabular}{|c ||c | c|} 
 \hline
 %& BB & CV\\ [0.5ex] 
 \begin{minipage}[b]{0.85\textwidth} \end{minipage} & BB & CV\\
 %\multicolumn{1}{|m{\tempwidth}||}{} &  BB & CV\\
 \hline\hline
 VP2 & 1212.82 mV & 1095.91 mV\\ 
 B2 & 771.68 mV & 895.59 mV\\  
 P3 & 744.92 mV & 626.80 mV\\  
 B3 & 882.29 mV & 1009.31 mV\\  
 P4 & 883.65 mV & 789.99 mV\\  
 B4 & 968.59 mV & 1060.28 mV\\  
 P5 & 642.24 mV & 581.91 mV\\ 
 B5 & 1135.03 mV & 1170.99 mV\\ 
 P6 & 670.47 mV & 622.13 mV\\ 
 B6 & 763.57 mV & 692.92 mV\\ [1ex] 
 \hline
\end{tabular}
\end{table}

\subsection{\label{supp:subsec:noise}Considerations on noise sources}

In the case of BB shuttling we identify several noise sources that can conceivably lead to shuttling fidelities below the limit set by dephasing in a static dot, in other words to consider phase-flip mechanims that are intrinsically connected to the tunneling events. First, in case the charge transitions are not perfectly adiabatic with respect to the tunnel couplings, the uncertainty in the moment of charge transfer leads to dephasing, as the dots have a different Zeeman splitting. 
However, given the measured tunnel couplings (Supplementary Fig. ~\ref{supp:fig:tunnel_couplings}), the pulse ramp times used, and the simulations in Supplementary Fig. \ref{supp:fig:BB_sim}, it is unlikely that diabatic transitions will dominate the shuttling performance in the absence of noise. This is further substantiated by the fact that an increasing ramp time lowers the shuttling fidelity, as shown in Figure \ref{fig:fig2}f: in case diabatic transitions were caused by the pulse flanks, the reverse trend would have been observed. However, it is known that high-frequency noise can also cause diabatic transitions~\cite{krzywda_adiabatic_2020}. In this case, slower ramps will give noise more time to cause diabatic transitions and the experimentally observed trend agrees with the predicted one.

A next possible mechanism is spin-flip tunneling, which can occur due to the intrinsic spin-orbit interaction (SOI) or an effective SOI from the micromagnet when the tunnel coupling is not large enough compared to the Zeeman splitting. Spin flips caused by tunneling are also visible in shuttling experiments that test how well the spin polarization is preserved. In Si/SiGe devices with a micromagnet, ~\cite{noiri_shuttling-based_2022} found a spin flip probability 14 times smaller than the phase flip probability, suggesting other phase flip processes are dominant during tunneling.

A plausible BB shuttling limitation, as stated in other works ~\cite{yoneda_coherent_2021, van_riggelen-doelman_coherent_2024}, is that charge noise strongly couples to the electron when shuttling through the zero detuning point of the double quantum dots. Given the large Zeeman splitting differences between the two sites of each DQD, the resonance frequency is highly sensitive to detuning fluctuations close to zero detuning (see Supplementary Fig. ~\ref{supp:fig:tunnel_couplings}), which leads to enhanced dephasing. The BB shuttling simulations in Supplementary Section \hyperref[supp:subsec:BB_sim]{C} and the experimentally observed dependence of the phase-flip probability on the ramp time corroborates this notion.

Lastly, hyperfine noise can cause spin dephasing. In this case a motional narrowing effect due to shuttling could possibly increase the coherence time, which has been experimentally observed in GaAs \cite{mortemousque_enhanced_2021} and in $^{nat}$Si/SiGe \cite{struck_spin-epr-pair_2024}. In this work, no noticeable motional narrowing was detected in bucket-brigade shuttling, though it is worth noting that the present device has only 0.08\%  magnetic nuclei in the quantum well versus 4.67\% for natural silicon devices and 100\% in GaAs.

\subsection{\label{supp:subsec:BB_sim}Simulation of dephasing during bucket brigade shuttling}

\begin{figure*}[ht]
\centering
\includegraphics[width=0.9\textwidth]{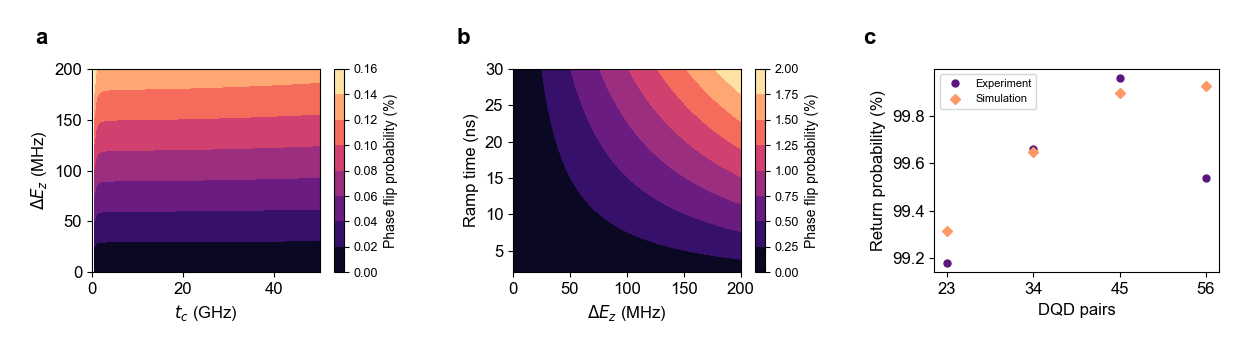}
\caption{\label{supp:fig:BB_sim} \textbf{Dephasing during BB shuttling.} a) Contour plot showing the phase flip probability as a function of the tunnel coupling $t_{c}$ and the Zeeman splitting difference $\Delta E_{z}$ between sites of a double quantum dot. b) Contour plot showing phase flip probability as a function of $\Delta E_{z}$ and the ramp time. c) Comparison of experimental and simulated return probabilities for different double quantum dot pairs. The parameters used for the simulation are extracted from the data in Supplementary Fig.~\ref{supp:fig:tunnel_couplings}}
\end{figure*}

In this section, we simulate the dephasing during the bucket-brigade shuttling process in the double quantum dots. Following \cite{benito_electric-field_2019}, we model the charge and spin dynamics in DQDs using the Hamiltonian
\begin{equation}
H_{s} = \frac{\varepsilon}{2}\tau_{z} + t_c\tau_{x} + \frac{1}{2} g\mu_{B} \Bigl[ B_z \sigma_{z} +(b_x\sigma_{x}+b_z\sigma_{z})\tau_{z} \Bigr]
\end{equation}
Here, $\tau_{x,y,z}$ and $\sigma_{x,y,z}$ correspond to the Pauli matrices in the charge and spin sector, respectively. $\varepsilon$ represents the detuning between the two dots, while $t_{c}$ denotes the tunnel coupling between DQDs. 2$b_{x}$ and 2$b_{z}$ indicate the differences in the x and z components of the magnetic field between the DQDs, induced by the micromagnet. $B_z$ indicates the external magnetic field (T).
When the inhomogeneous magnetic fields are weak, the transverse gradient has a second-order effect on the energy splitting of the spin qubit, while the longitudinal gradient introduces a first-order effect. As a result, the Zeeman splitting $E_s$ can be approximated as
\begin{equation}
E_s \simeq E_z-\frac{E_z^2-\varepsilon^2}{2 E_z\left(\Omega^2-E_z^2\right)}\left(g \mu_B b_x\right)^2-\frac{\varepsilon}{\Omega} g \mu_B b_z,
\end{equation}
where $\Omega=\sqrt{\varepsilon^2+4 t_c^2}$, and $E_z=g\mu_{B}B_z$.

Assuming detuning noise which fluctuates on timescales smaller than $E_s$, and assuming furthermore a small noise amplitude $\delta_{\varepsilon}$ ($\left| \frac{\partial E_s}{\partial \varepsilon} \delta_{\varepsilon} \right| \ll E_s$), 
we can estimate the dephasing rate $\Gamma_s$ for the spin qubit by employing time-independent perturbation theory, as
\begin{equation}
\begin{aligned}
\Gamma_{s} & =\left[\operatorname{Var}\left(\frac{\partial E_s}{\partial \varepsilon} \delta_{\varepsilon}+\frac{1}{2} \frac{\partial^2 E_s}{\partial \varepsilon^2} \delta_{\varepsilon}^2\right) / 2\right]^{1 / 2} \\
& =\left[\gamma_{s}^{(1)^2}+\gamma_{s}^{(2)^2}\right]^{1 / 2},
\end{aligned}
\end{equation}
where $\gamma_{s}^{(1)}=\gamma_\varepsilon \frac{\partial_{\varepsilon} E_s}{\partial \varepsilon}, \gamma_{s}^{(2)}=\gamma_\varepsilon ^2 \frac{\partial_{\varepsilon}^2 E_s}{\partial \varepsilon^2}$, and $\gamma_\varepsilon =\sigma_{\varepsilon} / \sqrt{2}$, where $\sigma_{\varepsilon}$ is the standard deviation of the fluctuations $\delta_{\varepsilon}$.

Assuming a completely adiabatic shuttling process, the phase flip probability due to detuning noise accumulated during shuttling can then be calculated as
\begin{equation}
1 - F_{p} = 1 - \exp \bigg( \int_{0}^{t_{\text{r}}} - \Gamma_{s}(t)t \, dt \bigg) \,
\label{eq:phaseflip}
\end{equation}
where $t_{\text{r}}$ is the detuning ramp time. We note that this expression assumes that the noise fluctuates during the shuttling rounds, as can be expected from motional narrowing effects~\cite{struck_spin-epr-pair_2024}. This is also consistent with the experimentally observed exponential decay of spin coherence in Ramsey-style shuttling experiments. 
%${\varepsilon}(t)$ from the left dot $\varepsilon _l$ to the right dot $\varepsilon _r$.

Supplementary Fig. ~\ref{supp:fig:BB_sim}a shows the simulated phase flip probability per shuttle hop as a function of $t_{c}$ and $b_{z}$, calculated with $E_z=$ \SI{48.8}{\micro eV}, $g \mu_B b_x=$ \SI{0}{\micro eV}, and $t_{r}=$ \SI{2}{\nano s}.

Supplementary Fig.~\ref{supp:fig:BB_sim}b shows the same simulation but as a function of tunnel coupling $t_{c}$ and ramp time $t_{r}$, calculated with $E_z=$ \SI{48.4}{\micro eV}, $g \mu_B b_x=$ \SI{0}{\micro eV}, and $t_{c}=$ \SI{124.2}{\micro eV}. Similar to the experimental data shown in Figure ~\ref{fig:fig2}f, the phase flip probability increases as the ramp time increases.

Supplementary Fig. ~\ref{supp:fig:BB_sim}c shows a comparison of the return probability obtained from the experiment and the simulation. For the latter, the tunnel couplings and Zeeman splitting differences are extracted from Supplementary Fig. ~\ref{supp:fig:tunnel_couplings}. Furthermore, $\gamma_{\varepsilon}=$ \SI{6.5}{\micro eV} is used, based on the average $T_{2}^{*}$ and $1/\Gamma_{s}$, taking into account the difference in measurement time.
Except for dot 5-6, the phase flip probability tends to decrease as $b_{z}$ decreases in both simulation and experiment.

\subsection{\label{supp:subsec:tunnelcoupling} Lever arms and tunnel couplings in BB}

\begin{figure*}[ht]
\centering
\includegraphics[width=0.82\textwidth]{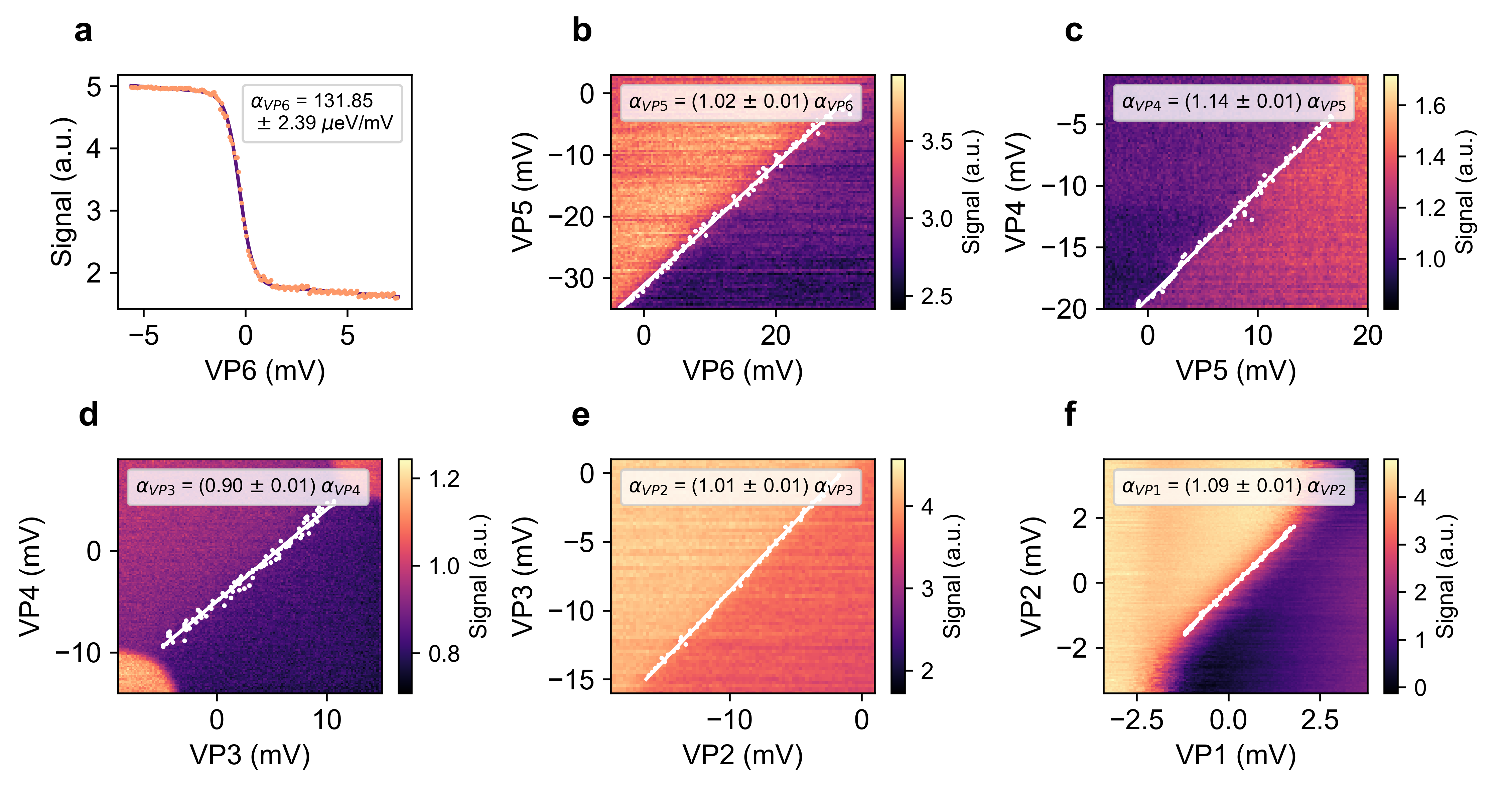}
\caption{\label{supp:fig:leverarm} {\bf Lever arm analysis.} a) Charge transition for dot 6. b)-f) Charge stability diagram near the interdot transition, showing a charge detector signal  as a function of two virtual gate voltages. The transition is extracted for each horizontal scan (white points) and the ratio of the lever arms of the two virtual gates  is estimated from the slope of the transition. Data taken at 500 mK.
}
\end{figure*}

For bucket-brigade shuttling, the tunnel couplings between the dots are crucial for the shuttling performance. Here we estimate the lever arms of the virtual plunger gates and use them to extract the tunnel couplings between the quantum dots. Supplementary Fig. ~\ref{supp:fig:leverarm}a shows the polarization line of the first electron in dot 6 at a mixing chamber temperature of \SI{500}{\milli \K}. The barrier to the reservoir is sufficiently closed such that the transition is mostly temperature broadened. This allows us to extract the lever arm $\alpha_{VP6}$ of virtual gate VP6 by  fitting the transition line to  $a \, {VP6} + b + c \, \Big( 1 + e^ {\, \alpha_{VP6} \, (VP6 - d\,) \, \beta_e} \Big)^{-1}$. Here, $a$, $b$, $c$ and $d$ are fitting parameters and $\beta_e = \big( k_B \, T_e \big) ^{-1}$ with $k_B$ the Boltzmann constant and $T_e$ the electron temperature, set to \SI{500}{\milli \K}. From the slope of the respective (1,0)-(0,1) anticrossings, we extract the relative lever arms of the virtual plunger gates in each double quantum dot. This yields a lever arm value for each virtual plunger gate.

In Supplementary Fig. ~\ref{supp:fig:tunnel_couplings}, the tunnel coupling $t_c$ between each quantum dot is obtained by fitting the spin resonance frequency at the interdot transition to $a \, \big( \epsilon - \epsilon_0 \big) + b + c\, \frac{\epsilon - \epsilon_0}{ \sqrt{ \big( \epsilon - \epsilon_0 \big)^2 + 4\, t_c^2}}$, where $\epsilon$ is the double dot detuning and $a$, $b$, $c$, $\epsilon_0$ and $t_c$ are fitting parameters. This corresponds to a simplified picture ~\cite{feng_control_2023}, where we assumed that the tunnel coupling is large compared to the Zeeman splitting, and that spin-dependent tunneling and the Stark shift difference between the dots are negligible.

\begin{figure*}[ht]
\centering
\includegraphics[width=0.95\textwidth]{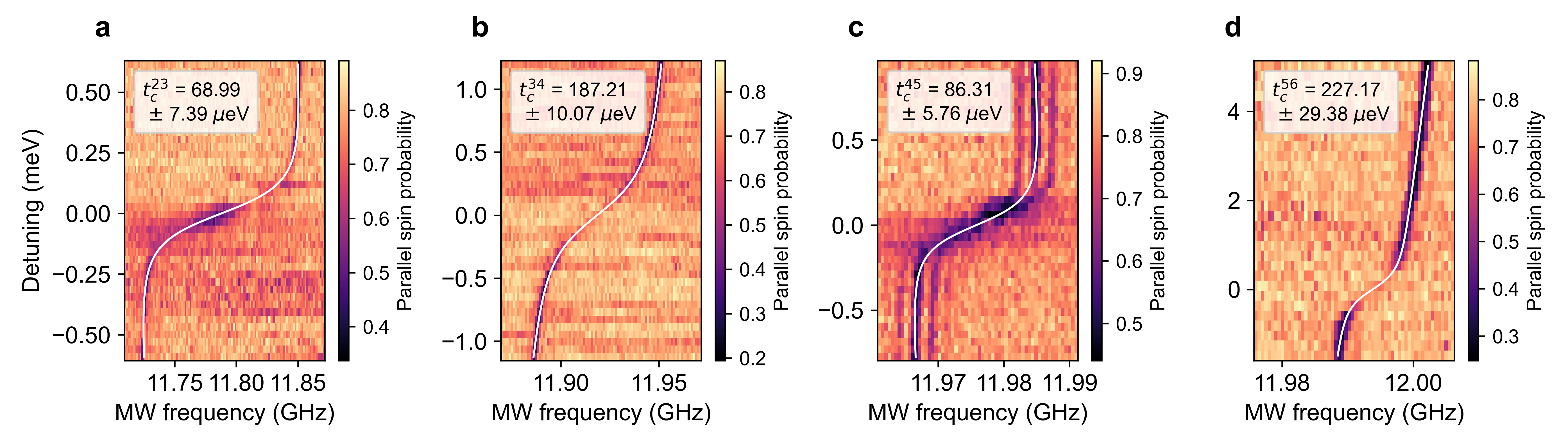}
\caption{\label{supp:fig:tunnel_couplings}{\bf Fitting the tunnel couplings.} Measured spin probability as a function of applied microwave frequency as the interdot detuning is scanned across the interdot transition for each double quantum dot. From the transition, the interdot tunnel coupling is extracted (see text).
}
\end{figure*}

\subsection{\label{supp:subsec:valley} Valley splitting}

Magnetospectroscopy measurements of the two-electron singlet-triplet energy splittings $E_{ST}$ in this device were performed and reported by  ~\cite{degli_esposti_low_2024}, albeit in a different cooldown and gate voltage configuration than these shuttling experiments. The estimated $E_{ST}$ values for all quantum dots are indicated in Table \ref{Supp:tab:valley}. $E_{ST}$ is a lower bound for the single-particle valley splitting $E_{v}$ in strongly confined quantum dots ~\cite{ercan_strong_2021} and is also the relevant metric for the size of the Pauli-spin-blockade readout window.
For our shuttling experiments, in each quantum dot the singlet-triplet energy splitting is significantly larger than the Zeeman splitting. Dephasing at spin valley hot-spots is therefore suppressed.

\begin{table}[h]
\centering
\caption{$E_{ST}$, as reported by ~\cite{degli_esposti_low_2024}, for each quantum dot in this device, which serves as a lower bound for the single-particle valley splitting.}
\label{Supp:tab:valley}
\begin{tabular}{||c c c c c c||} 
 \hline
 QD 1 & QD 2 & QD 3 & QD 4 & QD 5 & QD 6 \\ [0.5ex] 
 \hline\hline
 \SI{208}{\micro eV} & \SI{174}{\micro eV} & \SI{276}{\micro eV} & \SI{208}{\micro eV} & \SI{243}{\micro eV} & \SI{278}{\micro eV}\\  [1ex] 
 \hline
\end{tabular}
\end{table}

\subsection{\label{supp:subsec:CV_shuttle_coherence} Dephasing time during conveyor shuttling}

\begin{figure}[H]
\centering
\includegraphics[width=0.45\textwidth]{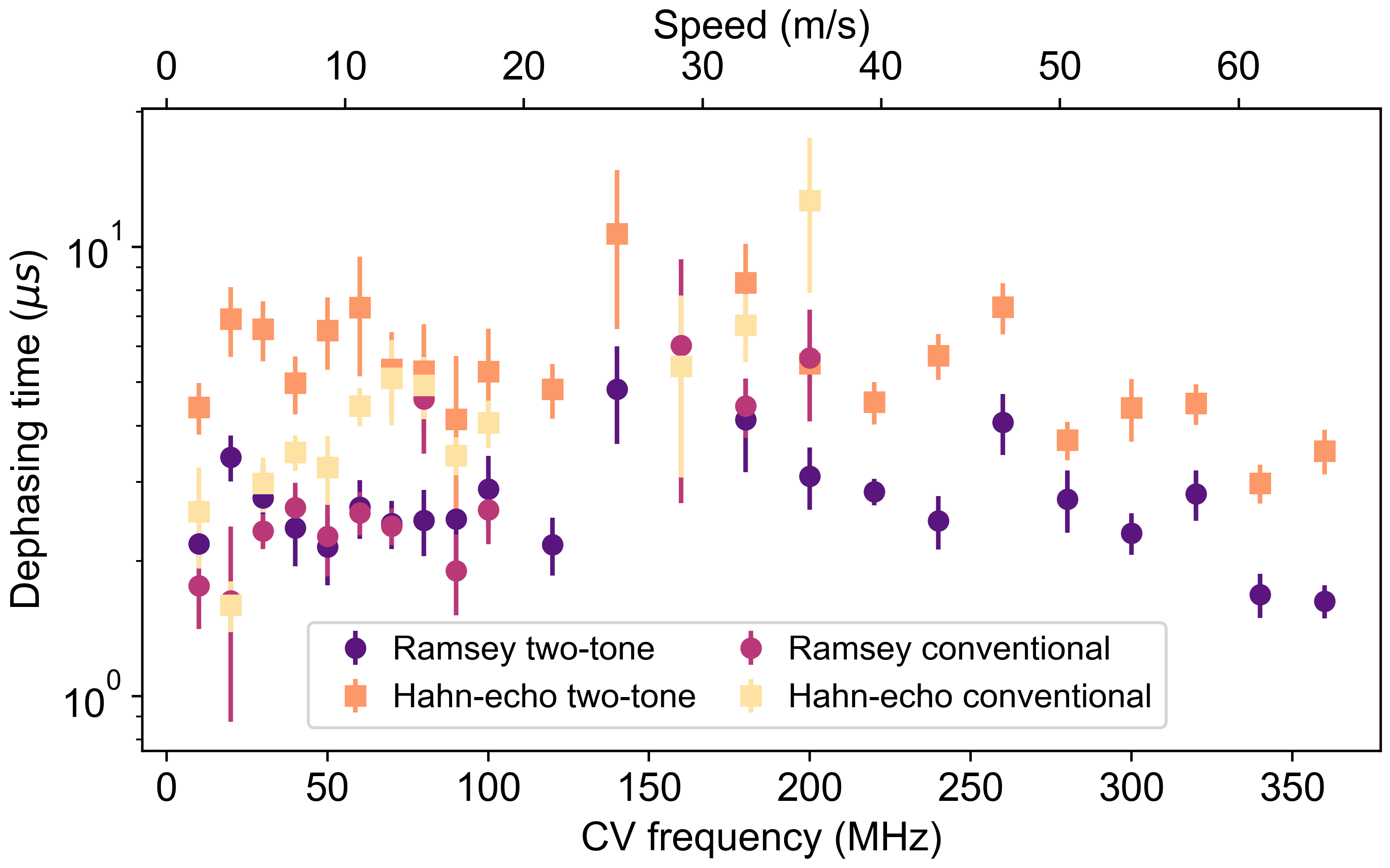}
\caption{\label{supp:fig:CV_T2}{\bf Coherence during conveyor shuttling.} Dephasing time and Hahn-echo decay time in a moving conveyor, both for the conventional and two-tone conveyor.
}
\end{figure}

The data represented in Figures ~\ref{fig:fig3}d and ~\ref{fig:fig3}e show the phase flip probabilities for  conventional and two-tone conveyor-mode shuttling per interdot distance $d$. As the transfer speed is varied, the total duration between initialization and readout is not identical for each da\textbf{}ta point. To investigate the effect of the conveyor shuttling process itself, we can fit the dephasing time of the spin during shuttling. Supplementary Fig. ~\ref{supp:fig:CV_T2} shows this dephasing time when conveyor-mode shuttling during the wait times of Ramsey and Hahn-echo sequences. For conveyor frequencies between 10 and 100 MHz, the average dephasing time during shuttling is \SI{2.49}{\micro \s} for the conventional, and \SI{2.69}{\micro \s} for the two-tone conveyor. In the case of a static conveyor potential shown in Figure ~\ref{fig:fig1}d, the dephasing time is \SI{1.45}{\micro \s} on average. This increase in dephasing time can likely be explained by motional narrowing. We therefore conclude that, different from the bucket-brigade results, the motion of the conveyor-mode potential does not limit the transfer fidelity, and probably enhances spin coherence.

\subsection{\label{supp:subsec:CV_loss} Data gap for the 100-160 MHz conveyor frequency band.}

Here we comment on the missing data points in Fig. ~\ref{fig:fig3}d for conveyor frequencies between 100 MHz and 160 MHz. In this frequency range, we do not observe Ramsey or Hahn-echo oscillations for any number of conveyor shuttle rounds. An arguably similar loss of charge shuttle fidelity at specific conveyor frequencies was observed in ~\cite{seidler_conveyor-mode_2022}, where the authors speculate the origin to be a resonance with a charge defect. We note that in our device the RF reflectometry circuits have resonances located at 135 MHz and 142 MHz. A conveyor operating at these frequencies could therefore lead to a loss of readout. However, we observe that the frequency band where shuttling fails is much broader than the RF resonances and that two-tone conveyor shuttling is not affected.

\subsection{\label{supp:subsec:device} Device fabrication}
    
\begin{figure}[H]
\centering
\includegraphics[width=0.42\textwidth]{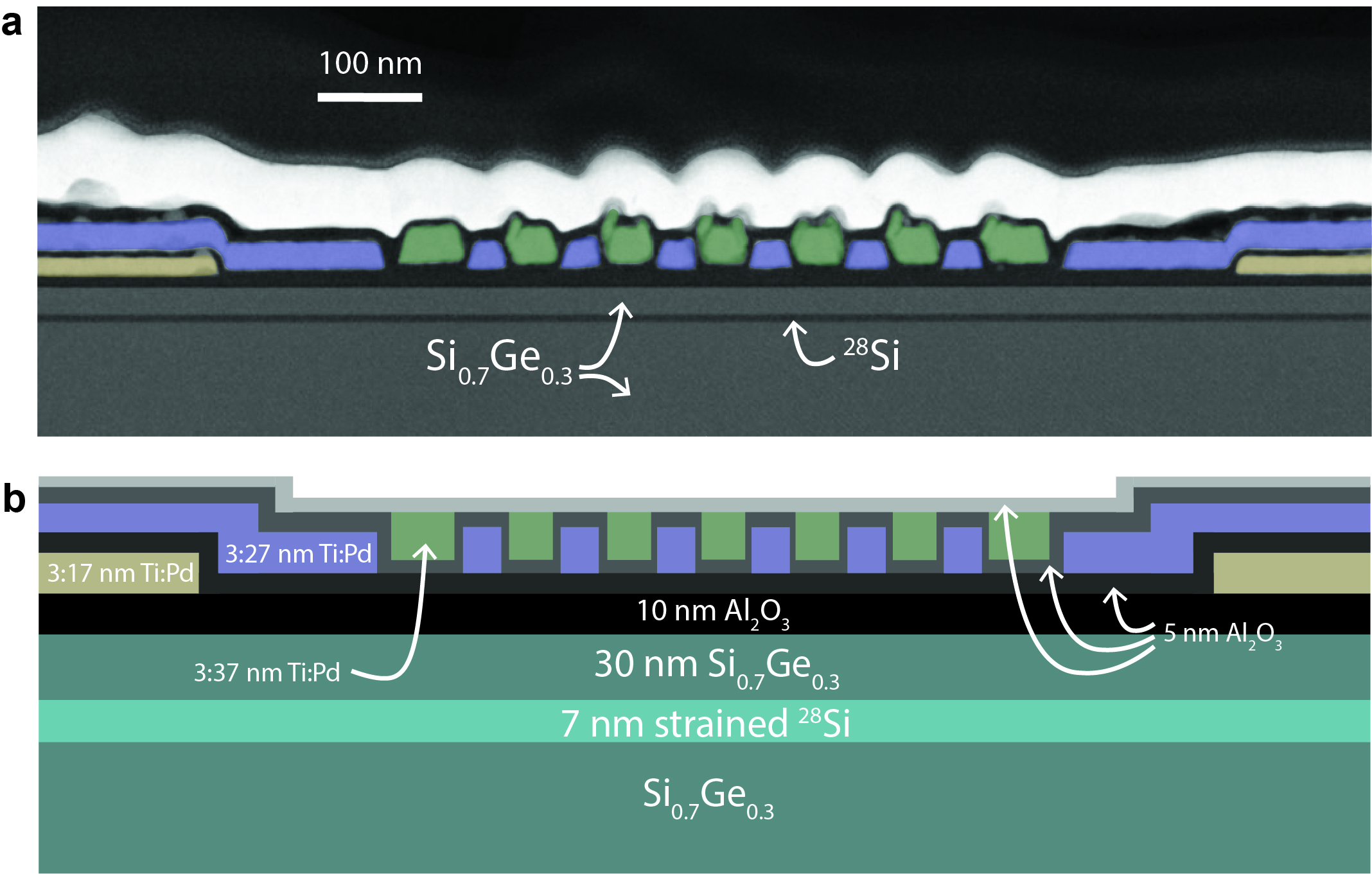}
\caption{\label{supp:fig:TEM}{Cross-section of a nominally identical device before deposition of the cobalt micromagnet. a) False-colored transmission electron microscope (TEM) cross-section image of the active area of the device. Eight (sensing) dot plunger (blue), seven barrier (green) and two screening (yellow) gates are visible. The white material on top of the device is a Pt cap added for improved imaging. Light grey areas in the \ch{Al2O3} do not correspond to electrical shorts between metallization layers, but are caused by local topography that is averaged across the thickness of the TEM lamella. b) Schematic representation of the cross-section, also indicating the different oxide layers.
}}
\end{figure}

    The device used in this work is fabricated on a \ch{^{28}Si}/SiGe heterostructure ~\cite{lawrie_quantum_2020}. Figure ~\ref{supp:fig:TEM} depicts a cross-section of the active area of the device. First, a \SI{1.5}{\micro \meter} linearly graded \ch{Si_{1-x}Ge_x} buffer is grown on a Si wafer. On top of that, a relaxed \SI{300}{\nano \meter} thick \ch{Si_{0.7}Ge_{0.3}} spacer is grown, following by a \SI{7}{\nano \meter} isotopically purified (800 ppm) tensile-strained \ch{^{28}Si} quantum well (QW) ~\cite{degli_esposti_low_2024}. Another \SI{30}{\nano \meter} \ch{Si_{0.7}Ge_{0.3}} spacer passivated with dichlorosilane at 500 $^{\circ}$C ~\cite{degli_esposti_wafer-scale_2022} separates the QW from the gate stack. Ohmic contacts to the two dimensional electron gas in the QW are made using phosphorus-ion implantation. A \SI{10}{\nano \meter} \ch{Al2O3} layer precedes three layers of Ti:Pd deposited using electron beam evaporation. The Ti:Pd gate layers have a thickness of 3:17, 3:27, 3:37 \SI{}{\nano \meter}, respectively, and are separated by \SI{5}{\nano \meter} thick \ch{Al2O3} layers deposited by atomic layer deposition. Finally, another \SI{5}{\nano \meter} thick \ch{Al2O3} layer is deposited on top of the gate stack, followed by a 5:200 \SI{}{\nano \meter} thick Ti:Co micromagnet, used for addressing and driving of the qubits.

%\bf

\subsection{\label{supp:subsec:BB6} Bucket-brigade shuttling from quantum dot 2 to 6}

\begin{figure}[H]
\centering
\includegraphics[width=0.4\textwidth]{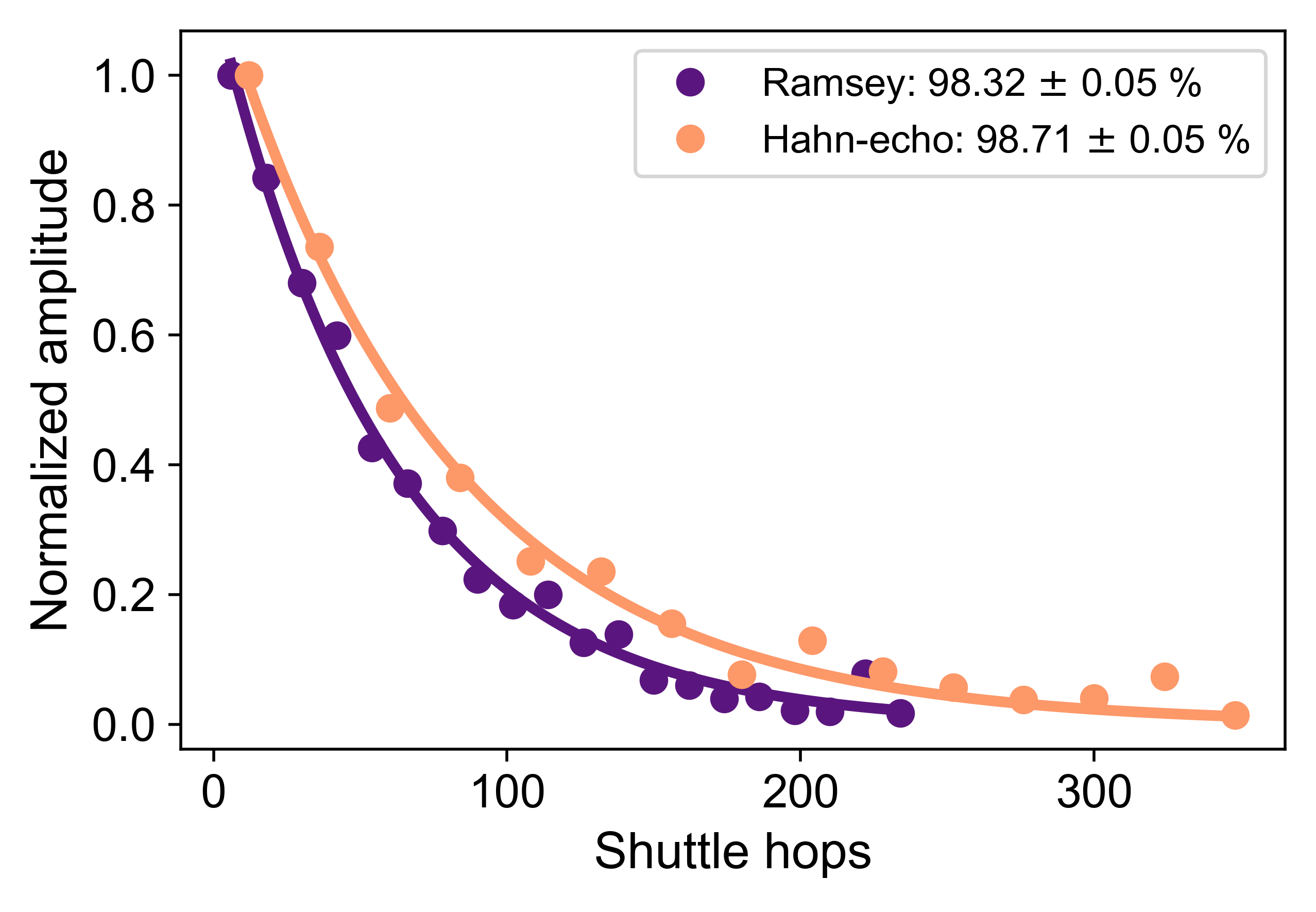}
\caption{\label{supp:fig:BB_Q2Q6} {\bf BB shuttling between quantum dot 2 and 6.} Normalized amplitudes of Ramsey- and Hahn-echo-style measurements during bucket-brigade shuttling between QD2 and QD6.
}
\end{figure}

In Supplementary Fig. \ref{supp:fig:BB_Q2Q6} the decay of Ramsey and Hahn-echo fringes during BB shuttling between dot 2 and dot 6 is presented. Bucket brigade shuttling across the entire array is therefore possible. Nonetheless, including dot 6 in the BB chain considerably decreases the shuttling performance. In Supplementary Fig. ~\ref{supp:fig:Polarization} the spin polarization as a function of the number of shuttle hops is shown. The relaxation rate is very small when the spin is shuttled between quantum dot 2 and 5. When shuttling between quantum dot 2 and 6, the relaxation rate increases dramatically. We have furthermore observed a low lever arm of gate B5 and a small charging energy in dot 5. We speculate that the small orbital energy could induce diabatic charge excitations. Given the artificial spin-orbit interaction from the micromagnet, this would affect not only the spin dephasing rate (phase flips) but also the spin relaxation rate (spin flips) during shuttling. We point out that a comparison of Supplementary Fig. ~\ref{supp:fig:Polarization} to the dot 5-6 results in Fig. ~\ref{fig:fig2}b and ~\ref{fig:fig2}c is difficult. Not only does the former include barrier pulses during shuttling, the VB5 voltage operation point during the 5-6 detuning pulse is 20 mV higher compared to the double dot shuttling in Fig. ~\ref{fig:fig2}.

\begin{figure}[H]
\centering
\includegraphics[width=0.4\textwidth]{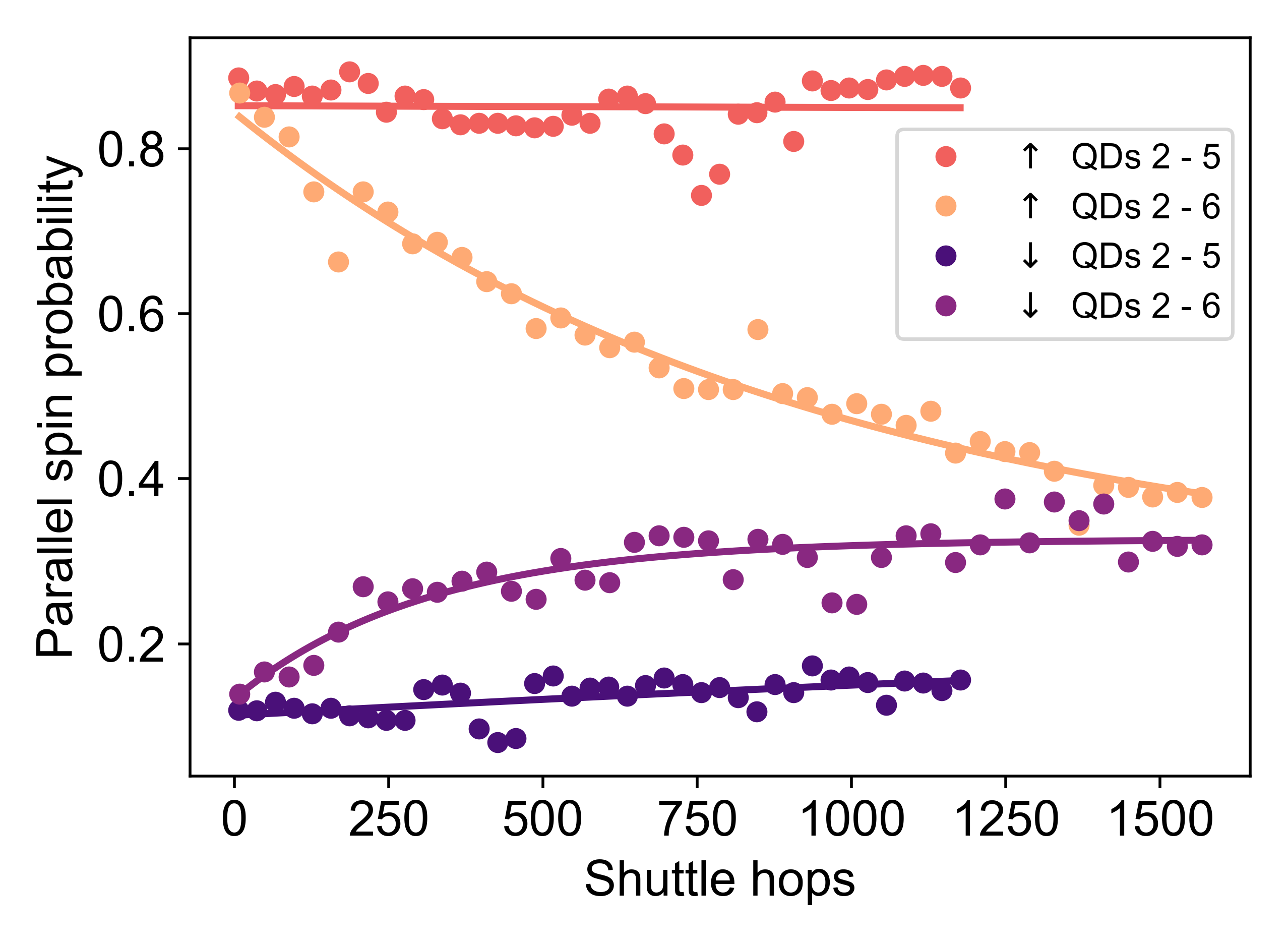}
\caption{\label{supp:fig:Polarization} {\bf Spin flips when shuttling in BB mode.} Spin relaxation during bucket-brigade shuttling between dot 2 and 5 and between dot 2 and 6. Including quantum dot 6 in the shuttling trajectory leads to a significant increase in the relaxation rate. The solid lines are exponential fits to the data.
}
\end{figure}

\subsection{\label{supp:subsec:BB_idling} Idling site for double dot bucket-brigade shuttling}

The idling operation in Fig. ~\ref{fig:fig2}a is performed in different dots for the Ramsey and Hahn-echo sequences. Ideally, the idle time would be equally divided between the two sites in order to also eliminate the effect of any $T_{2}^{*}$ difference in the dots. To evaluate its impact, we assume dephasing with Gaussian decay in both sites of the double quantum dot and simulate the difference in the phase-flip probability when performing the final idling time in different dots. This difference decreases linearly with the number of shuttle hops, as the idling time necessarily becomes shorter. From our simulation, the difference is limited to -0.04\%, -0.02\% and 0.02\% per hop for pair 2-3, 3-4 and 4-5, respectively, which falls within the standard deviation to the fit in Fig. ~\ref{fig:fig2}b. The negative (positive) sign of the difference, refers to a slight over(under)estimation of the phase-flip probability. We can therefore conclude that, despite the non-ideal implementation of the idling time in ~\ref{fig:fig2}a, there is no significant effect on the extracted phase-flip probabilities.

\subsection{\label{supp:subsec:driven_magnet} Two-tone conveyor shuttling fidelity with driven external electromagnet}

\begin{figure}[h]
\centering
\includegraphics[width=0.45\textwidth]{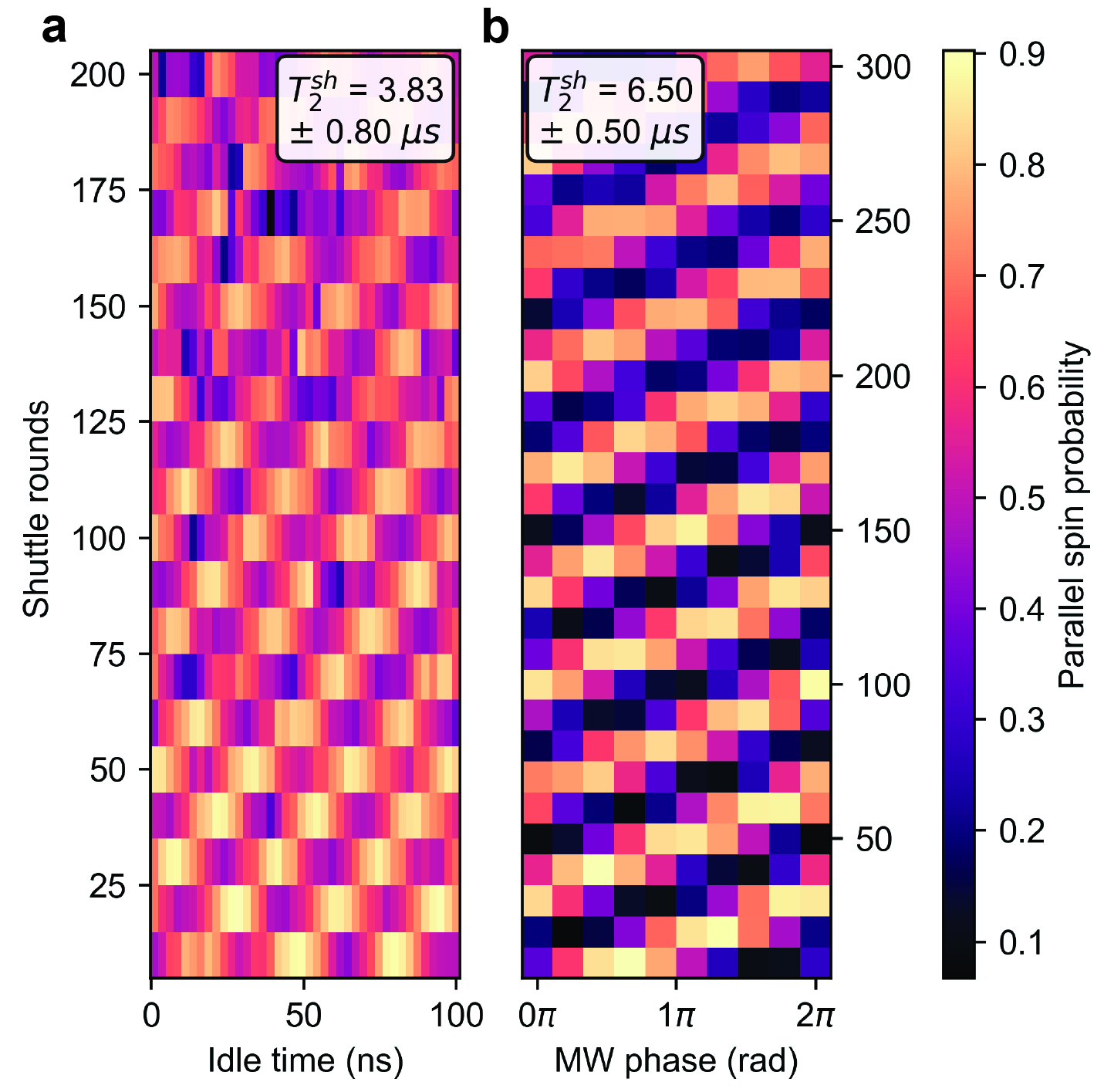}
\caption{\label{supp:fig:shuttlerounds} {\bf Coherence time depends on the operating mode of the superconducting magnet.} A single electron is shuttled back-and-forth using a two-tone conveyor at a main frequency of 300 MHz and with a 4 ns one-way shuttling time. a) Ramsey-like measurement, as used for ~\ref{fig:fig3}d, where the idling time refers to a variable wait time after shuttling a number of rounds and before readout. The superconducting magnet is in driven mode. b) Similar Ramsey-like measurement, where the idling time is replaced with a virtual Z gate with variable phase before the final X$_{90}$. In this case, the superconducting magnet is in persistent mode. The insets indicate an estimated coherence time during shuttling, extracted from fitting the decay in oscillation amplitude with shuttle rounds using an exponential function.}
\end{figure}

After completion of the shuttling measurements, we found that, for a spin in a static conveyor, up to twice longer $T_{2}^*$ values are obtained with the superconducting magnet in persistent mode, reducing external field fluctuations. This suggests that global magnetic-field fluctuations were likely affecting the spin-shuttling fidelities. This is confirmed by Supplementary Fig. ~\ref{supp:fig:shuttlerounds}, where a significant increase in the coherence time during conveyor-mode shuttling is observed when the state of the magnet is persistent. For comparison, we show shuttling IRB with the magnet in driven mode in Supplementary Fig. ~\ref{supp:fig:IRB_driven}. The shuttling frequency and time per round are identical, though the shuttling gate comprises 23 instead of 24 shuttle rounds.

\begin{figure}[H]
\centering
\includegraphics[width=0.395\textwidth]{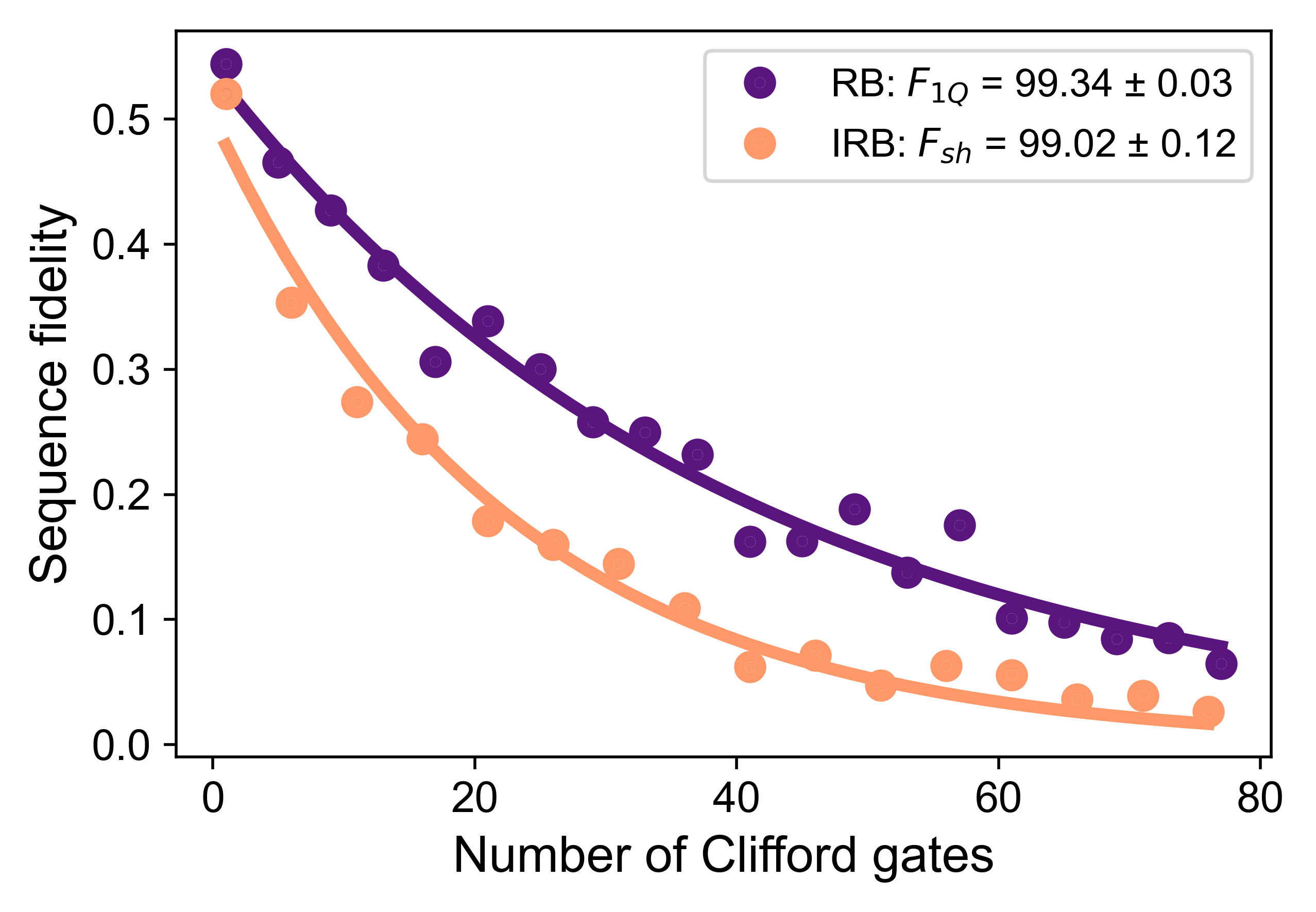}
\caption{\label{supp:fig:IRB_driven} {\bf Conveyor shuttling IRB with the superconducting magnet in driven mode.} Interleaved randomized benchmarking with 23 shuttle rounds, yielding a shuttle fidelity of 99\% for an effective \SI{10}{\micro \meter} shuttling distance.
}
\end{figure}

\begin{figure}[H]
\centering
\includegraphics[width=0.395\textwidth]{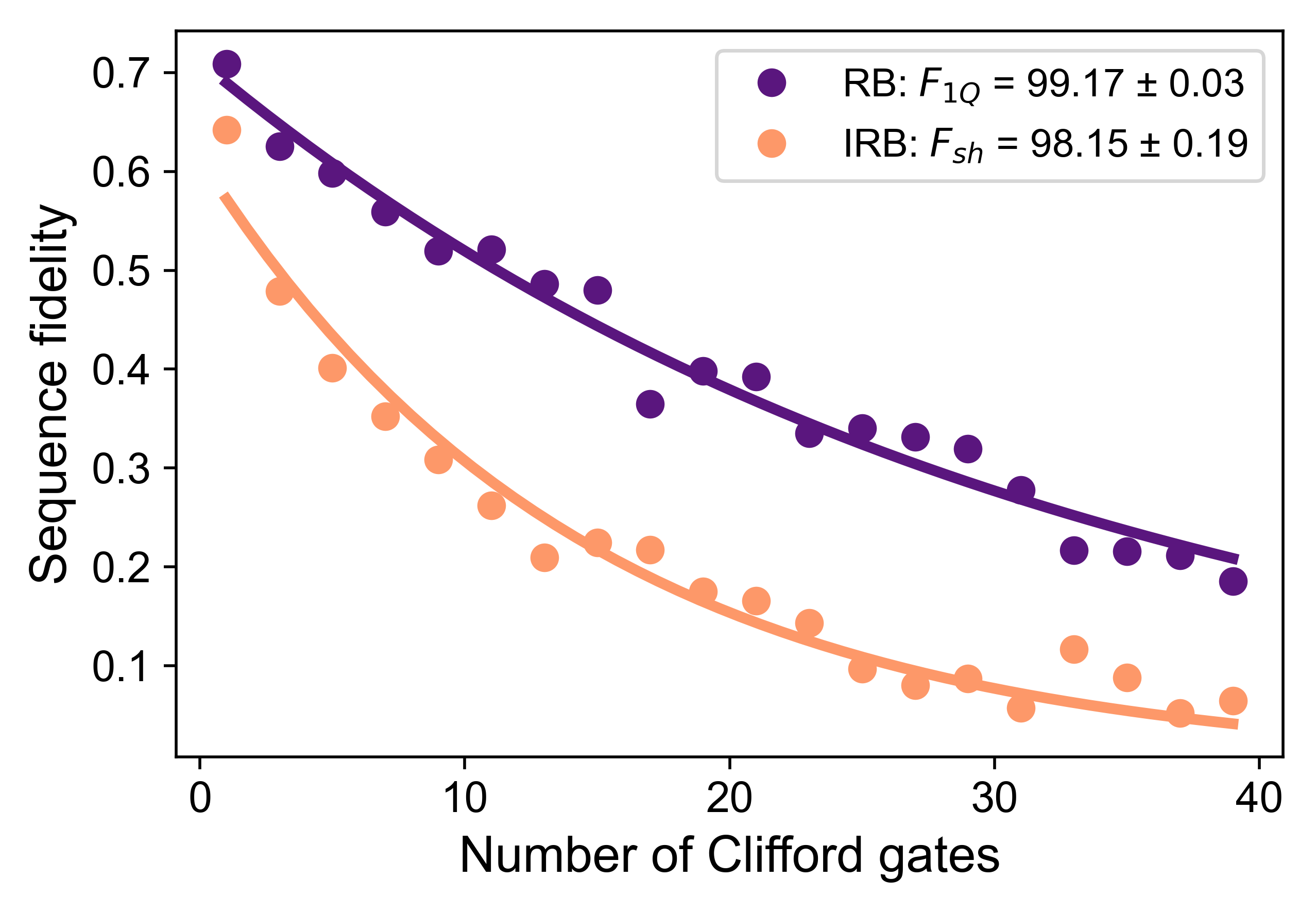}
\caption{\label{supp:fig:IRB_equal} {\bf Conveyor shuttling IRB with equal DC voltages.} Interleaved randomized benchmarking with equal DC voltages applied to all the channel gates in the same layer. Here, the shuttling gate comprises a single shuttle round and a phase gate.
}
\end{figure}

\subsection{\label{supp:subsec:IRB_equal} Two-tone conveyor shuttling fidelity with equal DC voltages}

In Figure ~\ref{fig:fig3}c we demonstrate that conveyor-mode shuttling with equal DC offsets is possible, though requiring higher pulse amplitudes. Here we show interleaved randomized benchmarking of shuttling using a two-tone conveyor at 80 MHz while emulating a lack of individual gate control. To do so, the conveyor amplitude is set to 90 mV for each gate, regardless of the gate layer. This is with the exception of VP2 (70 mV), as needed to keep the reference electron confined in dot 1. We apply pulse offsets such that the barrier gates B2, B3, B4 and B5 in the conveyor are set to a DC voltage of 900 mV, and the plunger gates P3, P4, P5 and P6 are set to a DC voltage of 710 mV while shuttling. Gates on the edges of the device were exempted, as they are crucial not only for the initialization and readout but also for the transition of the single electron into the conveyor. Moreover, it is imperative to avoid losing the shuttled electron to the reservoirs. In Supplementary Fig. ~\ref{supp:fig:IRB_equal} the fidelity for shuttling over a distance of 432 nm without individual gate control is determined to be 99.18 $\pm$ 0.19 \%. Extrapolating to \SI{10}{\micro \meter} from the case with individual DC control, this amounts to a fidelity of 65.08 $\pm$ 2.90 \%.

\begin{figure}[H]
\centering
\includegraphics[width=0.4\textwidth]{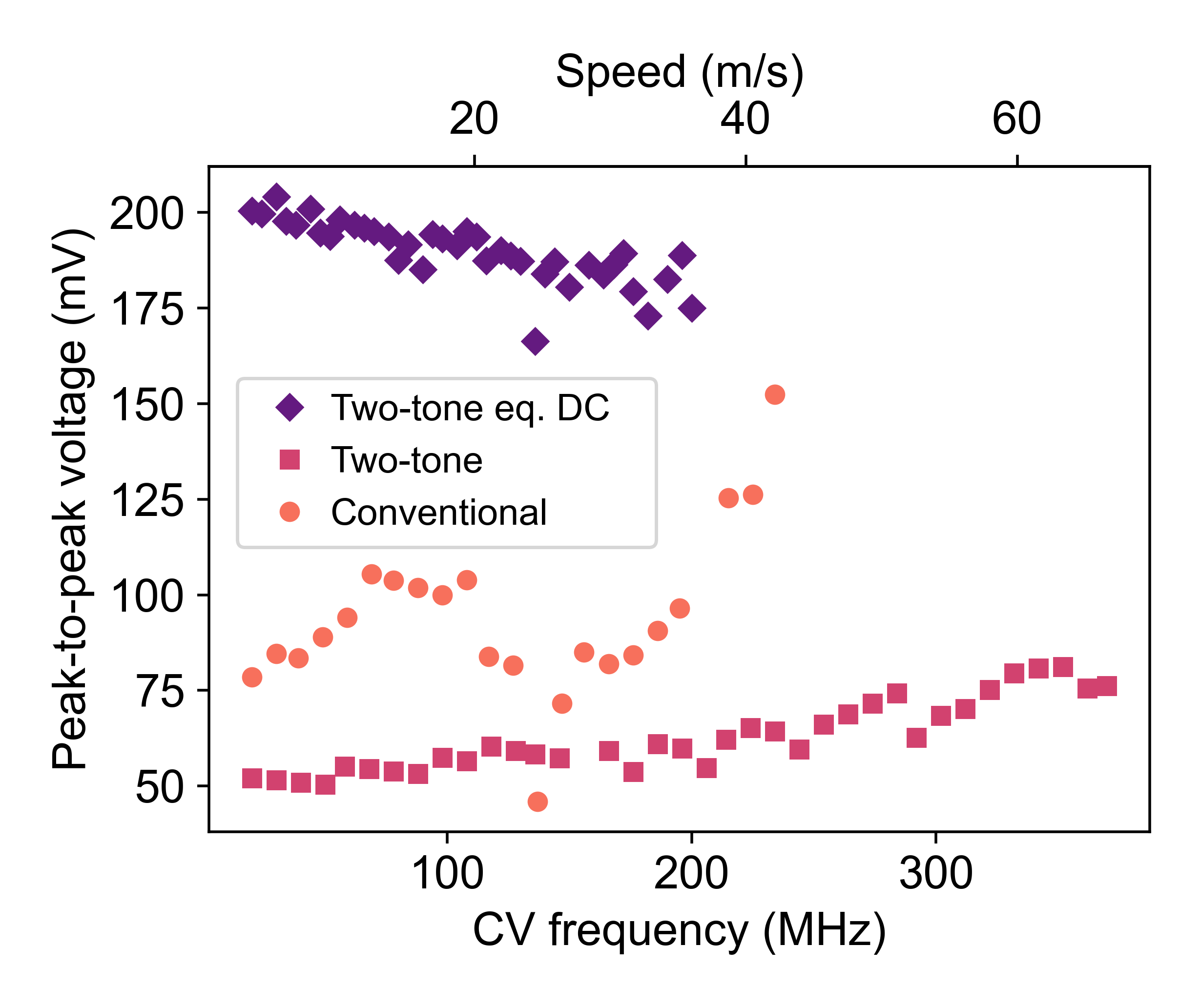}
\caption{\label{supp:fig:Amplitudes_corrected} {\bf Required peak-to-peak pulse voltage after AWG filter.} Minimal peak-to-peak gate voltage required for successful charge transfer by conveyor shuttling, accounting for the filter function built-in to the AWG. 
}
\end{figure}

\subsection{\label{supp:subsec:Amplitudes_corrected} Required conveyor amplitude corrected for AWG filter}

Whereas Figure ~\ref{fig:fig3}c takes into account the attenuation in the transmission lines, it does not consider the output filter of the waveform generator. In Supplementary Fig. ~\ref{supp:fig:Amplitudes_corrected} we correct for the phase and amplitude distortion of the applied sine signals using the measured AWG filter function. Note that in the two-tone case, the filter response is different for both sine waves, leading to an alteration of the shape of the traveling wave potential with increasing conveyor frequency. Therefore, here we indicate the peak-to-peak voltage of the applied signals. It is not clear to us why the required peak-to-peak voltage for successful charge transfer (slightly) decreases with conveyor frequency for the two-tone implementation with equal DC voltages. If anything, we were a priori expecting that higher amplitudes might be needed for higher conveyor frequencies, as seen in the other two cases shown. As stated in the main text, smaller amplitudes are required for the two-tone conveyor than for the conventional conveyor. This is what we were expecting given that the potential barriers surrounding a moving dot are wider for the two-tone conveyor (see also the diagrams in Fig.~\ref{fig:fig3}a,b), and an electron is therefore less likely to escape from the potential minimum in which it is meant to travel.

% \subsection*{\label{supp:subsec:micromagnet}Micromagnet simulation}

\begin{figure*}[hb]
\centering
\includegraphics[width=0.95\textwidth]{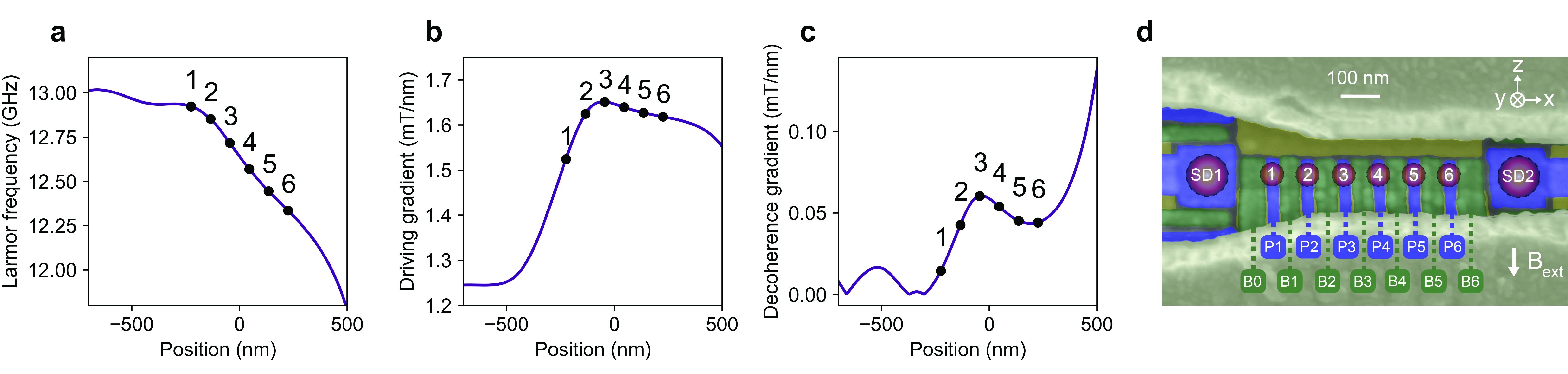}
\caption{\label{supp:fig:uMag}\textbf{Micromagnet simulation.} a) Simulated qubit frequencies along the array at an external magnetic field of 0.26 T, with the target qubit positions along the x direction indicated. The micromagnet is magnetized along the z direction with a magnetization vector of M = (0, 0, -1.5) T. The experimentally observed Larmor frequency trend in Figure ~\ref{fig:fig1}c, though also monotonic, is opposite to the simulated one. Simulations with reasonably altered parameter values of the magnet geometry can produce a parabolic-like trend, similar to experimental observations in \cite{philips_universal_2022}, though a Larmor frequency trend akin to the experimental one is not found. Nonetheless, the experimentally observed frequency trend has been measured consistently in multiple samples. b) Simulated transverse driving gradient of the micromagnet c) Simulated decoherence gradient of the micromagnet. All simulations were performed using the python package magpylib. d) False-colored SEM image of a nominally identical device, indicating the frame of reference and the applied magnetic field direction.
}
\end{figure*}

\begin{figure*}[hb]
\centering
\includegraphics[width=\textwidth]
{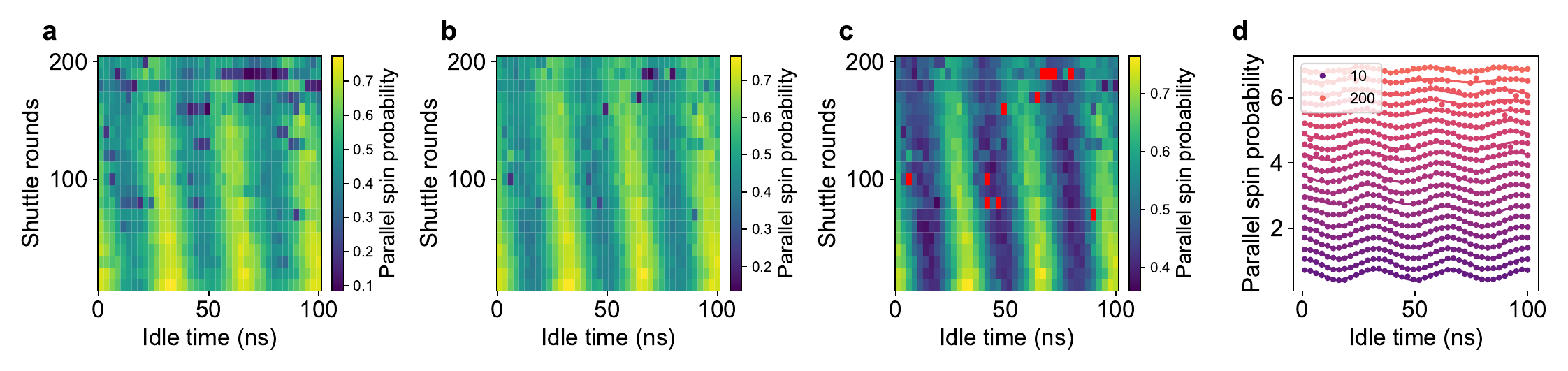}
\caption{\label{fig:supp_fig:analysis} 
\textbf{Data analysis for Fig. 3d of the main text.} a) Raw data example of a Hahn-echo-like measurement used in Fig. 3d of the main text while shuttling a number of rounds (back-and-forth from approximately under gate P2 to under gate P5) with a two-tone conveyor having a main frequency of 280 MHz. b) The Hahn-echo-like measurement from a) after dynamical rethresholding by double gaussian fitting for each point in the 2D plot. c) The Hahn-echo-like measurement from b) after setting a global minimum probability to remove the outliers (red) that were not recovered with dynamical rethresholding. d) Fitting the parallel spin probability of the data in c) with a cosine function for each number of shuttle rounds in order to extract the decay in amplitude. Each line is offset by 1/3 for clarity.
}
\end{figure*}

% \subsection*{\label{supp:subsec:setup}Measurement set-up}
\begin{figure*}[h]
\includegraphics[width=0.78\textwidth]{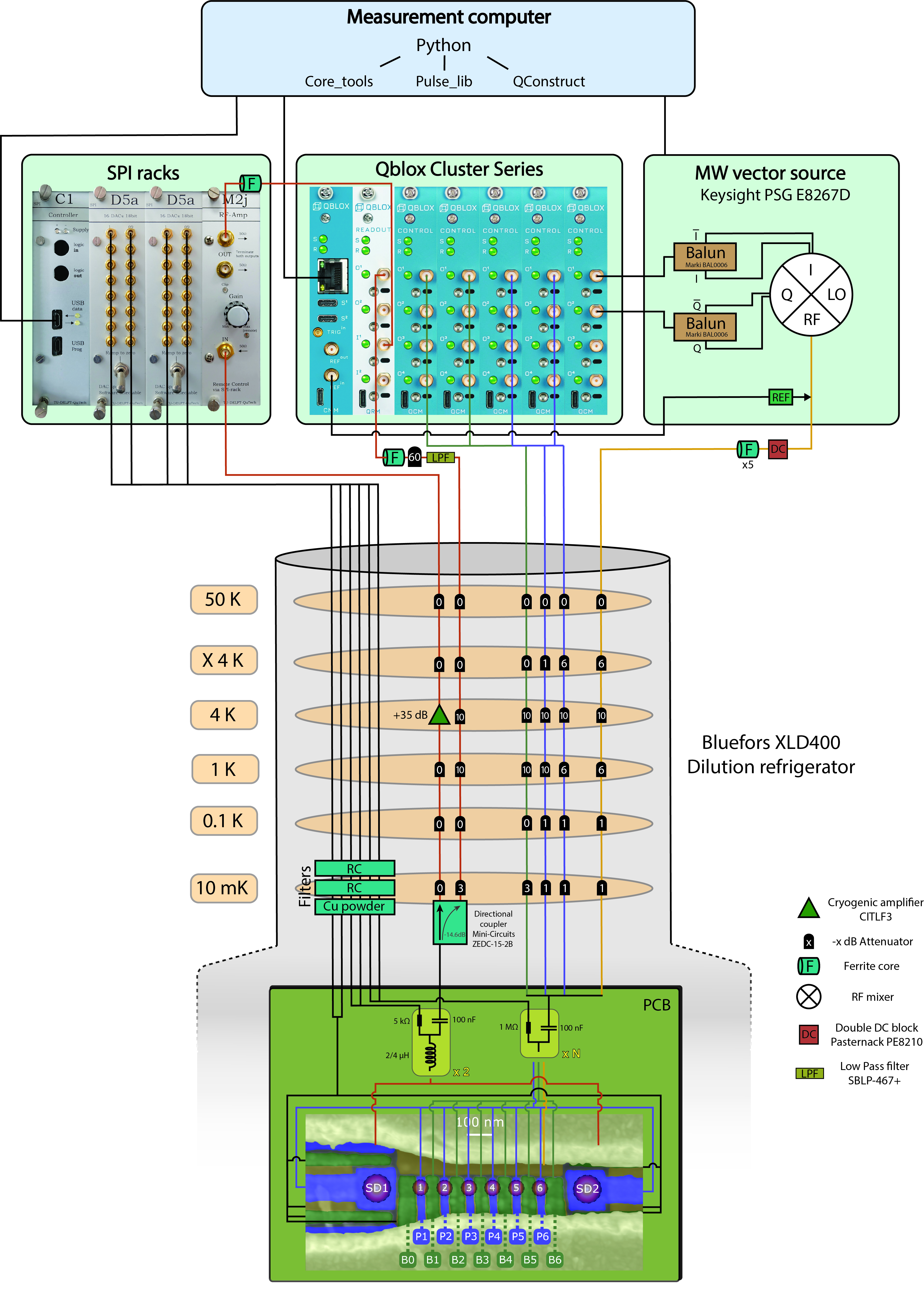}
\caption{\label{fig:supp_fig:setup} 
\textbf{Schematic of the experimental setup.} The experiments were executed using three main pieces of electronics (shown in green). The SPI rack, powered by batteries and equipped with gyrators, houses two sets of in-house built DACs. They provide low-noise DC voltages that pass through RC and copper powder filters, before arriving at the PCB. Baseband pulses below 400 MHz are applied using QCM modules in the Qblox Cluster Series. The pulses are attenuated at different stages in the Bluefors XLD400 dilution refrigerator, amounting to -23 or -24 dB attenuation in total. Bias tees on the PCB with an RC time constant of 100 ms combine the pulses with the DC voltages. All plunger and barrier gates in the channel, as well as the sensing dot plungers, are equipped with these bias tees. For readout, RF reflectometry is implemented on both sensing dots. RF signals generated by the Qblox QRM module are sent through a ferrite core, a -60 dB attenuator and a low pass filter at room temperature. After -23 dB additional attenuation in the dilution refrigerator, the RF signal is split in two and directed to a bias tee in series with an off-chip superconducting NbTiN inductor for each sensing dot. The inductor is wire-bonded directly to a reservoir accumulation gate neighboring the sensing dot. The reflected RF signal passes through the directional coupler and is amplified at the 4 K stage using a Cosmic Microwave Technology CITLF3 cryogenic amplifier. It is then amplified again at room temperature using an in-house built M2j low-noise amplifier, after which IQ demodulation is performed by the QRM. We use a Keysight PSG E8267D microwave vector source for EDSR driving. The I and Q input signals are generated by a Qblox QCM and differentiated using Marki BAL006 Baluns. Ferrite cores and a double DC block are added to the microwave line. After a total of -24 dB attenuation in the refrigerator, the MW signal is routed via a coplanar waveguide on the chip to the top screening gate.
}
\end{figure*}
\pagebreak

\end{document}